\documentclass[fleqn,usenatbib]{mnras}

\usepackage{newtxtext,newtxmath}

\usepackage[T1]{fontenc}

\DeclareRobustCommand{\VAN}[3]{#2}
\let\VANthebibliography\thebibliography
\def\thebibliography{\DeclareRobustCommand{\VAN}[3]{##3}\VANthebibliography}

\usepackage{graphicx}
\usepackage{amsmath}

\usepackage[inline]{enumitem}
\usepackage{amsmath}
\renewcommand{\vec}[1]{{\mathbf{#1}}}
\usepackage{multirow}
\usepackage{comment}
\usepackage{placeins}

\title[Hot Jupiters and stellar clustering]{On hot Jupiters and stellar clustering: the role of host star demographics}

\author[M. V. Kontiainen et al.]{
Mika V. Kontiainen,$^{1}$\thanks{E-mail: mvk26@cam.ac.uk (MVK)}
Cathie. J. Clarke,$^{1}$
and Andrew J. Winter$^{2}$
\\
$^{1}$Institute of Astronomy, University of Cambridge, Madingley Road, CB3 0HA, UK\\
$^{2}$Astronomy Unit, Department of Physics and Astronomy, Queen Mary University of London, Mile End Road, London E1 4NS, UK
}

\date{Accepted XXX. Received YYY; in original form ZZZ}

\pubyear{\the\year{}}

\begin{document}
\label{firstpage}
\pagerange{\pageref{firstpage}--\pageref{lastpage}}
\maketitle

\begin{abstract}
The variation in hot Jupiter (HJ) occurrence across stellar environments holds clues as to the dominant formation channels of these extreme planets. Recent studies suggest HJ hosts preferentially reside in regions of high phase space density, possibly reflecting natal environmental conditions. These regions are kinematically cold ($|v| < 40\:{\rm km\:s^{-1}}$), prompting the alternative hypothesis that the correlation reflects an age bias: planetary systems in overdensities are systematically younger and therefore less likely to have undergone tidal inspiral and destruction. We test whether the apparent excess of HJs in phase space overdensities arises from differences in intrinsic host properties -- mass, metallicity, age -- which may correlate with phase space density or whether there is evidence for an additional environmental effect. We derive homogeneous estimates for the mass, metallicity, and age of planet-hosting stars using 2MASS and \textit{Gaia} DR3 photometry, parallaxes, and self-consistent spectroscopic and spectrophotometric observables. In a sample of 2265 confirmed exoplanet hosts, we find a significant relative excess of HJs orbiting stars in overdense regions. However, we also find that overdensities preferentially host younger, more massive, and more metal-rich stars compared to underdensities. After correcting for these differences, either by detrending the phase space density against age or by matching host properties across subsamples, we find no significant differences in the HJ populations between over- and underdense regions. Our results suggest that the previously reported correlation between HJ occurrence and phase space density is driven by underlying differences in host star demographics rather than an intrinsic environmental effect.
\end{abstract}

\begin{keywords}
stars: kinematics and dynamics -- planetary systems -- exoplanets
\end{keywords}

\section{Introduction}

\noindent Hot Jupiters (HJs) are gas giant exoplanets with orbital periods of less than $\sim$ 10 days, occupying an extreme region of parameter space that challenges conventional planet formation theories. The close proximity of HJs to their host stars raises questions about their origins, as the standard core accretion model predicts that giant planets should form beyond the water-ice line, where solid material is more abundant \citep{helled_giant_2014}. Understanding how HJs reach their present-day orbits remains an open question, with several potentially complementary formation pathways proposed.

The three main mechanisms proposed for HJ formation are:
\begin{enumerate*}[label=(\roman*)]
    \item in-situ formation,
    \item primordial disc migration, and
    \item high-eccentricity migration
\end{enumerate*}
\citep{dawson_origins_2018}. Correlations between the occurrence rate of HJs and their host and environmental properties can shed light on which of the formation processes is dominant. In particular, it has been suggested that high-eccentricity migration could be enhanced in clustered environments, where frequent stellar encounters can dynamically perturb cold Jupiters (CJs) onto highly eccentric orbits \citep{zakamska_excitation_2004, rodet_correlation_2021}. These planets may then undergo tidal circularisation, leading to the formation of a hot Jupiter \citep{brucalassi_search_2016}.

By defining a 6D phase space density metric and decomposing the local stellar neighbourhoods of planet-hosting stars into over- and underdensities, \citet{winter_stellar_2020} found that hot Jupiter hosts preferentially reside in phase space overdensities. This was interpreted as reflecting the formation of hot Jupiters in clustered environments, the kinematic signatures of which remain discernible in phase space over Gyr timescales while spatial structures are washed out \citep{kamdar_dynamical_2019}. However, \citet{mustill_hot_2022} questioned this finding, pointing out that the phase space density is strongly correlated with the peculiar velocity of the host star. As stars are thought to form with low peculiar velocities and undergo dynamical heating via interactions over time \citep{wielen_diffusion_1977, freeman_galactic_1987, nordstrom_geneva-copenhagen_2004, tarricq_3d_2021}, the abundance of hot Jupiters in kinematically cold overdensities and their corresponding dearth in hot underdensities would therefore reflect a decline in HJ occurrence rate with stellar age.

Giant planet occurrence is known to correlate with both stellar metallicity \citep{fischer_planet-metallicity_2005, petigura_california-kepler_2018, osborn_investigating_2020} and mass \citep{johnson_giant_2010}. Recently, multiple authors have demonstrated that hot Jupiter hosts are on average younger and kinematically colder than stars hosting cold Jupiters or non-giant planets \citep{miyazaki_evidence_2023, chen_evolution_2023, swastik_age_2024, banerjee_host-star_2024}. Multiple explanations have been put forward for the paucity of hot giant planets around older stars, including the insufficient enrichment of the ISM for giant planet formation earlier in the history of the Galaxy \citep{Swastik_galactic_2022, boettner_populating_2024} and the tidal inspiral and eventual destruction of hot Jupiters over time \citep{hamer_hot_2019, tejada_arevalo_further_2021}.

While \citet{winter_stellar_2020} controlled for differences in host ages by imposing limits on the stellar age distributions, their analysis relied on heterogeneously derived values, restricting the sample to systems with age estimates available on the NASA Exoplanet Archive. Using homogeneously derived isochronal ages, \citet{adibekyan_stellar_2021} found hosts in phase space underdensities to be notably older than those in overdensities. However, as their analysis was limited to planets detected by the radial velocity (RV) method, the sample size ($N_\star = 58$) was insufficient to decisively confirm or rule out differences in planetary properties between the two subsamples. Whether the anti-correlation between HJ occurrence and stellar age can solely explain the correlation with stellar clustering identified by \citet{winter_stellar_2020} therefore remains an open question. Here we attempt to answer this question using a larger and more representative sample of planet-hosting stars for which we derive homogeneous age estimates.

In Section \ref{section:methods}, we begin by introducing our sample and key methodology. In Section \ref{section:results}, we present our results from splitting the sample by phase space density and correcting for differences in host properties. In Section \ref{section:discussion}, we discuss our findings and implications for the role of the stellar environment in shaping planet demographics. We conclude in Section \ref{section:conclusions} by summarising our results.

\section{Methods}
\label{section:methods}

\subsection{Host Sample}

\noindent We obtained the observed properties of all 5787 confirmed planets, corresponding to 4325 host stars, in the Planetary Systems Composite Data catalogue from the NASA Exoplanet Archive.\footnote{\href{https://exoplanetarchive.ipac.caltech.edu/}{exoplanetarchive.ipac.caltech.edu/} accessed in November 2024} Following \citet{winter_stellar_2020}, only systems with a reported stellar mass in the range 0.7--2.0~${\rm M_\odot}$ were retained, reducing the sample to 4659 planets and 3489 unique hosts. The host stars were then cross-matched with \textit{Gaia} DR3 sources with full \textit{Gaia} ($G$, $G_{\rm BP}$, $G_{\rm RP}$) photometry and 6D astrometry \citep{gaia_collaboration_gaia_2023}. We obtained unambiguous host matches for 3303 planets, corresponding to 2433 host stars. Finally, after filtering for planets with missing mass or semi-major axis values, we were left with 3092 planets and 2265 unique host stars. Adopting the \citet{winter_stellar_2020} definition of a hot Jupiter as a planet of mass $M_{\rm p} > 50~{\rm M_\oplus}$ and semi-major axis $a < 0.2$~AU, 570 stars in our sample are classified as HJ hosts. For the purposes of this paper, we refer to the complementary set of giant planets with $a \geq 0.2$~AU as cold Jupiters.

Rather than impose the 1.0--4.5~Gyr age cuts performed in previous studies based on the heterogeneously derived stellar ages from the NASA Exoplanet Archive, we set out to obtain homogeneous and mutually self-consistent age estimates for our sample. In order to obtain isochrone-based ages, we cross-matched our sample with the SWEET-Cat catalogue of planet-hosting stars \citep{santos_sweet-cat_2013, sousa_sweet-cat_2021}.\footnote{\href{https://sweetcat.iastro.pt/}{sweetcat.iastro.pt/}} Following \citet{adibekyan_stellar_2021}, we limited our sample to FGK stars, defined by \citet{Kunimoto_searching_2020} as having $T_{\rm eff}$ in the range 3900--7300~K. In making this cut, uncertainties in $T_{\rm eff}$ were propagated to ensure that only stars inconsistent with this range at the $1\sigma$ level were removed. We similarly removed evolved stars with $\log{g} < 4.0$ from the sample, resulting in a full set of stellar atmospheric parameters ($T_{\rm eff}$, $\log{g}$, [Fe/H]) for 1874 FGK main-sequence host stars, 711 of which were derived homogeneously using the same uniform methodology \citep{sousa_ares_2014}. While our main analysis is based on the homogeneous sample, we also show our results using the full heterogeneous catalogue collated from the literature in Appendix \ref{appendix:figures}.

The homogeneous SWEET-Cat parameters are largely derived from high-resolution ground-based spectra which are primarily available for stars hosting planets discovered by the radial velocity (RV) method. Close-in transit planets are therefore underrepresented relative to RV planets compared to the NASA Exoplanet Archive. To make up for this, we also queried the \citet[hereafter ZGR23]{zhang_parameters_2023} catalogue of stellar atmospheric parameters inferred from \textit{Gaia} XP spectra.\footnote{\href{https://doi.org/10.5281/zenodo.7692680}{doi.org/10.5281/zenodo.7692680}} We only selected sources for which the model confidence in all three inferred parameters ($T_{\rm eff}$, $\log{g}$, [Fe/H]) exceeds 0.5 by requiring \texttt{quality\_flags == 0}. Before applying any sample cuts, we inflated the reported stellar parameter uncertainties by adding a systematic error floor $\epsilon$ in quadrature. Following \citet{zhang_parameters_2023}, we chose $\epsilon = 100$~K in $T_{\rm eff}$ and 0.1~dex in $\log{g}$ and [Fe/H]. This resulted in a supplementary sample of 1826 FGK main-sequence host stars with full spectrophotometric data. Most of these are transit planet hosts, as large spectroscopic surveys, such as LAMOST \citep{zhao_lamost_2012} which \citet{zhang_parameters_2023} use as their training sample, are biased against nearby bright stars. The properties of the two samples used in our analysis are summarised in Table \ref{table:samples}.

\begin{table}
\centering
\begin{tabular}{l|c|c} 
\hline
 & SWEET-Cat & ZGR23 \\
\hline
Hosts (FGK MS) & 711 & 1826 \\
Hosts (0--13~Gyr) & 690 & 1392 \\
HJ hosts & 271 & 393 \\
\hline
Planets & 968 & 1896 \\
Transit planets & 492 & 1792 \\
RV planets & 471 & 86 \\
\hline
Source & Ground-based spectra & \textit{Gaia} XP spectra \\
Reference & \citet{santos_sweet-cat_2013} & \citet{zhang_parameters_2023} \\
\hline
\end{tabular}
\caption{Description of the homogeneous host samples used in this study.}
\label{table:samples}
\end{table}

\subsection{Isochrone Fitting}

\noindent To obtain isochronal ages for the host sample, we used the open-source code \texttt{isoclassify} \citep{Huber_asteroseismology_2017, berger_gaiakepler_2020, Berger_gaia-kepler-tess-host_2023} with the MIST isochrones \citep{Dotter_mesa_2016, Choi_mesa_2016} and the \texttt{Combined19} Galactic dust extinction map \citep{Drimmel_three-dimensional_2003, Marshall_modelling_2006, Green_3d_2019} from \texttt{mwdust} \citep{Bovy_galactic_2016}. As input variables we included the measured positions (RA, Dec), spectroscopic or spectrophotometric $T_{\rm eff}$, $\log{g}$, and [Fe/H] with their associated uncertainties as well as \textit{Gaia} DR3 parallaxes with uncertainties inflated by 30 per cent \citep{el-badry_million_2021}. For the ZGR23 sample, we used the revised \textit{Gaia} DR3 parallax values provided in the same catalogue and added a systematic uncertainty floor $\epsilon = 100$~K in $T_{\rm eff}$ and 0.1~dex in $\log{g}$ and [Fe/H] in quadrature, in line with the values used for sample selection. We also included photometry from \textit{Gaia} ($G$, $G_{\rm BP}$, $G_{\rm RP}$) and 2MASS ($J$, $H$, $K_{\rm s}$) \citep{skrutskie_two_2006} with a 0.01~mag floor uncertainty imposed across all bands.

\subsection{Phase Space Decomposition}

Following the procedure outlined in \citet{winter_stellar_2020}, we obtained the phase space density for each host star using the concept of Mahalanobis distance defined as
\begin{equation}
    d_{\rm M}(\vec{x}, \vec{y}) = \sqrt{(\vec{x}-\vec{y})^{\rm T} C^{-1} (\vec{x}-\vec{y})}
\end{equation}
where $\vec{x}$ and $\vec{y}$ are coordinate vectors in 6D phase space and $C$ is the covariance matrix calculated using the phase space positions (Galactocentric positions and velocities) of all stars within 80~pc of the host. To ensure the robustness of the metric, we limited our analysis to hosts with a minimum of 400 stars with full astrometry within 40~pc. For each host and up to 600 randomly drawn stars in the 40~pc neighbourhood, we calculated the Mahalanobis distances to all stars within 40~pc. For each of these, we then identified the 20th nearest neighbour based on the Mahalanobis distance and inverted this to obtain the local phase space density as $\rho_{\rm M,20} = 20 d_{\rm M,20}^{-6}$. We then normalised the phase space density of the host and its neighbours relative to the median in the neighbourhood.

Next, we decomposed the normalised $\log_{10}{\tilde\rho_{\rm M,20}}$ distribution using a two-component Gaussian mixture model. Before fitting the model, we removed outliers more than two standard deviations from the mean as well as those with $\tilde\rho_{\rm M,20} > 50$ as these are likely associated with gravitationally bound clusters. Where a one-component model produced a good fit ($p_{\rm null} > 0.05$), we attempted no further decomposition. Otherwise, we calculated the probability $P_{\rm high}$ of the host belonging to the high-density component of the Gaussian mixture model. Following \citet{winter_stellar_2020} and \citet{mustill_hot_2022}, we assigned hosts with $P_{\rm high} > 0.84$ to overdensities and those with $P_{\rm high} < 0.16$ to underdensities. Hosts with $\tilde\rho_{\rm M,20} > 50$ were assigned to overdensities regardless of $P_{\rm high}$.

As shown by \citet{kamdar_stars_2019}, proximity in both real and velocity space is necessary to identify stars that are likely conatal. In particular, by integrating stars on Galactic orbits over Gyr timescales, \citet{kamdar_spatial_2021} found that comoving pairs of stars within $\sim$ 40~pc of each other are more likely than not to be conatal. This justifies the choice of spatial volume constraint. However, as no constraint is imposed in velocity space, the spatial and velocity coordinates used in calculating $\tilde\rho_{\rm M,20}$ have differing dynamic ranges, such that velocity coordinates naturally dominate the metric. Whether a given host is designated as residing in a region of low or high phase space density is therefore determined to a greater extent by the kinematic coherence rather than the physical stellar density in the local 40~pc neighbourhood, as also observed by \citet{blaylock-squibbs_no_2024}. We discuss the implications of this distinction in greater detail in Section \ref{section:results}.

\section{Results}
\label{section:results}

\begin{figure}
\includegraphics[width=\columnwidth]{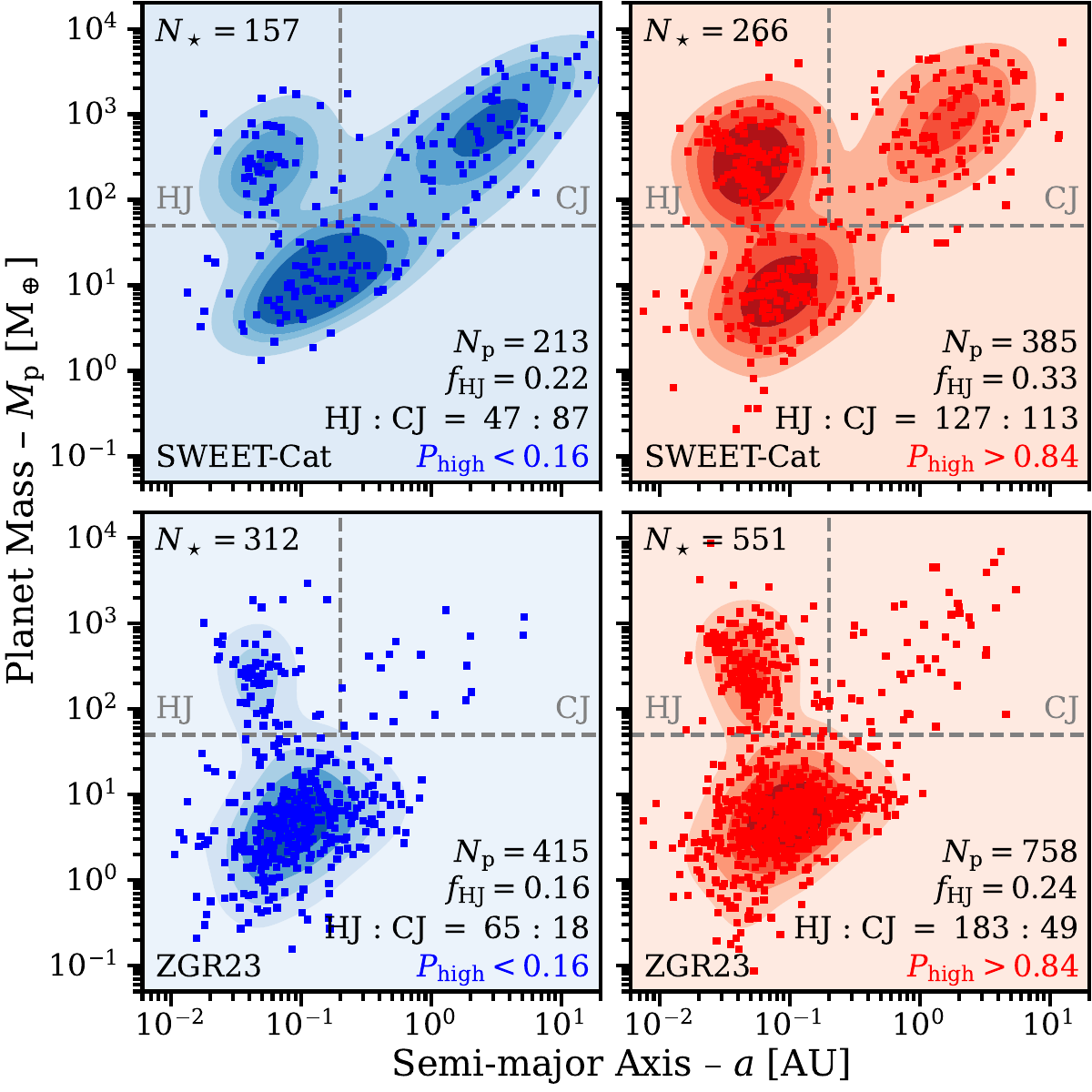}
\caption{The distributions of planet semi-major axes and masses in the homogeneous SWEET-Cat (top) and ZGR23 (bottom) host samples split by phase space density into underdensities (left) and overdensities (right) before matching for host properties or applying any age cuts. The region of the parameter space ($M_{\rm p} > 50~{\rm M_\oplus}$, $a < 0.2~$AU) occupied by hot Jupiters is indicated by dashed lines.
\label{fig:planet_properties}}
\end{figure}

\noindent Following the phase space decomposition but prior to applying any age cuts, 872 hosts are identified as residing in overdensities and 515 in underdensities. The remaining 878 hosts in our original sample cannot be assigned unambiguously to either component following the \citet{winter_stellar_2020} methodology. Of these, 91 reside in neighbourhoods with phase space density distributions well described by a single log-normal, while another 137 have fewer than 400 neighbours within 40~pc and are excluded from further analysis. The relative proportions of overdensity and underdensity stars in the SWEET-Cat and ZGR23 samples are consistent with the original sample. That more hosts are assigned to phase space overdensities than underdensities overall is to be expected given the time-dependent nature of phase space substructures and the observed tendency of younger stars to host more planets \citep{bashi_exoplanets_2022}.

The planet populations in the SWEET-Cat and ZGR23 samples before applying any age cuts are shown in Figure \ref{fig:planet_properties}. In both samples, hot Jupiters appear more abundant around overdensity stars with the overall fraction of hot Jupiters being 0.24--0.33 in overdensities and 0.16--0.22 in underdensities. To quantify the difference in the occurrence of hot Jupiters, we adopt a bimodal approach by assuming the number of events (HJ host) in a sample follows the Poisson distribution. Specifically we use the Poisson means test (E-test) to determine whether the number of stars that host hot Jupiters in each sample is significantly different. Before applying any age cuts, we find the difference in the number of HJ hosts to be moderately or highly significant with Poisson $p$-values of $2.4\times10^{-3}$ and $7.9\times10^{-4}$ in the SWEET-Cat and ZGR23 samples, respectively. Due to differences in the relative number of RV and transit planets, the hot-to-cold Jupiter ratio exhibits a notable difference (0.54 vs 1.1) between the two subsamples only in the SWEET-Cat sample, while the ratios in both ZGR23 subsamples are heavily skewed towards HJs (3.6 vs 3.7).

As more than a third of the original 2265 host stars are effectively discarded by the decomposition procedure, we also consider an alternative approach in which the normalised phase space density $\tilde\rho_{\rm M,20}$ is treated as a continuous variable. While the phase space densities of individual stars are strictly speaking only comparable within a given neighbourhood, this allows us to exploit the full sample, including hosts with ambiguous environment classifications. In line with more HJ hosts residing in overdensities after the phase space decomposition, we also find the distribution of normalised Mahalanobis phase space densities $\tilde\rho_{\rm M,20}$ to be significantly skewed towards higher values for HJ hosts. The median values of $\log_{10}{\tilde\rho_{\rm M,20}}$ for HJ and non-HJ hosts are 0.15 and $-0.06$ ($p_{\rm KS} \approx 2.4\times10^{-4}$) in SWEET-Cat and 0.17 and $-0.01$ ($p_{\rm KS} \approx 7.3\times10^{-6}$) in ZGR23.

While $\tilde\rho_{\rm M,20}$ is defined to quantify the degree of spatial clustering and kinematic coherence relative to the local neighbourhood, its physical interpretation is not entirely obvious. Motivated by this ambiguity, \citet{mustill_hot_2022} identified a strong correlation between the Mahalanobis phase space density $\tilde\rho_{\rm M,20}$ and the total peculiar velocity $|v|$ of the host star, implying that phase space density largely serves as a proxy for stellar kinematics. We similarly computed the Galactocentric velocities for all hosts in our sample by adopting the Sun's peculiar motion $(U,\:V,\:W) = (11.1,\:12.24,\:7.25)$~${\rm km\:s^{-1}}$ \citep{schonrich_local_2010} relative to a Local Standard of Rest (LSR) of (0, 235, 0)~${\rm km\:s^{-1}}$. Consistent with \citet{mustill_hot_2022}, we find hosts in phase space overdensities to be kinematically colder as measured by their total peculiar velocity $|v|$ relative to the LSR. Specifically, overdensity hosts have median peculiar velocity of 23.5~${\rm km\:s^{-1}}$ in both SWEET-Cat and ZGR23 samples, respectively, whereas the corresponding values for underdensity hosts are 65.1 and 66.6~${\rm km\:s^{-1}}$. The relationship between the Mahalanobis phase space density and total peculiar velocity is presented in Figure \ref{fig:rho_vpec}.

\begin{figure}
\includegraphics[width=\columnwidth]{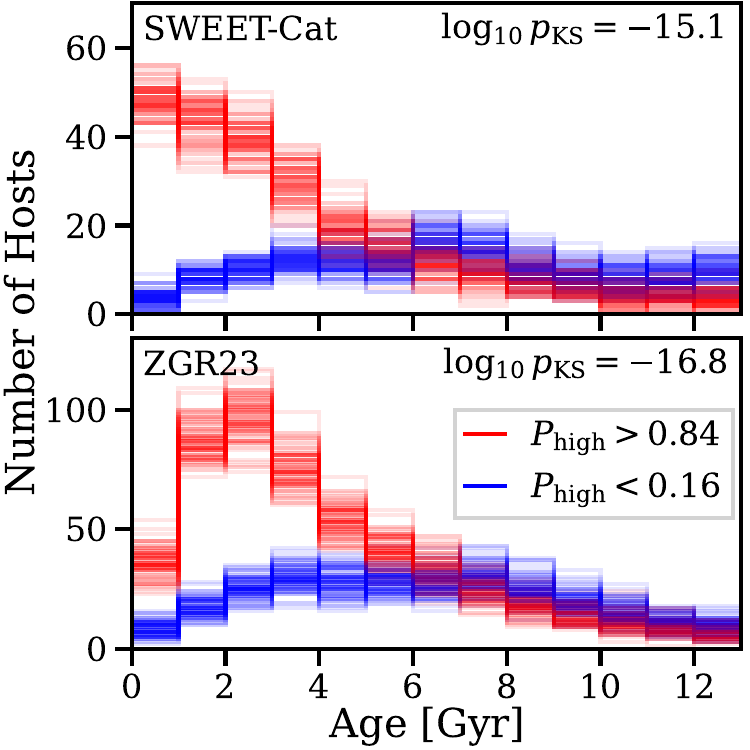}
\caption{Host star age distributions in the overdensity (red) and underdensity (blue) subsamples obtained using the homogeneously derived SWEET-Cat (top) and ZGR23 (bottom) stellar parameters. The faint lines show the distributions obtained by resampling the values from the corresponding \texttt{isoclassify} age posteriors 100 times.
\label{fig:sample_ages}}
\end{figure}

To determine the role of stellar age in driving the observed discrepancies, we obtained physically reasonable (0--13~Gyr) isochronal ages for 690 hosts in the homogeneous SWEET-Cat sample and 1392 hosts in the ZGR23 sample. The median relative uncertainties in the age estimates are 20 and 55 per cent, respectively. The stellar age distributions of the overdensity and underdensity samples, shown in Figure \ref{fig:sample_ages}, confirm that hosts in overdensities are significantly younger on average. In the SWEET-Cat sample, the median host age is 3.1~Gyr in overdensities and 7.4~Gyr in underdensities, while in the ZGR23 sample the corresponding values are 3.3~Gyr and 6.3~Gyr, respectively. The concurrent differences in stellar age and kinematics suggest that dynamical heating is at least partly responsible for the observed kinematic differences. We also find overdensity hosts to be more massive and more metal-rich on average, with the median values differing by approximately $0.1~{\rm M_\odot}$ and $0.06~{\rm dex}$. The latter difference is generally less pronounced, consistent with the relatively flat age–metallicity relation observed for Milky Way thin disc stars \citep{bensby_exploring_2014}.

As discussed in \citet{winter_stellar_2020} and \citet{adibekyan_stellar_2021}, not imposing an upper age limit can lead to a contamination of the underdensity sample by stars that were formerly part of phase space overdensities. A lower age limit is likewise required to ensure the orbits of massive planets have had time to stabilise \citep{bitsch_eccentricity_2020}. These considerations motivated \citet{winter_stellar_2020} to restrict their analysis to hosts with ages in the range 1--4.5~Gyr, which \citet{adibekyan_stellar_2021} further extended to 0.5--5~Gyr in order to preserve a larger sample size. On the other hand, if phase space substructures arise as a result of late-time Galactic-scale perturbations, as suggested by \citet{kruijssen_not_2021}, no strict limits on the host ages are required. In this study we remain agnostic as to whether high phase space density reflects initial conditions or is instead a consequence of stellar youth. However, to focus on the possibility that it traces formation conditions, we also explore restricting the sample to hosts in the 1--5~Gyr age range.

Similarly to \citet{adibekyan_stellar_2021}, we find that applying strict age cuts generally diminishes the difference in HJ occurrence between the two subsamples. In particular, we find the differences in the homogeneous SWEET-Cat sample to be sensitive to the adopted age range, with the previously significant excess of HJs in overdense regions largely disappearing ($p_{\rm Poisson} \approx 0.57$) when restricting the sample to the 301 stars with ages between 1 and 5~Gyr. The differences between the $\tilde\rho_{\rm M,20}$ distributions of HJ and non-HJ hosts are similarly no longer significant ($p_{\rm KS} \approx 0.21$), as also found by \citet{mustill_hot_2022}. While age estimates derived using stellar parameters from the full SWEET-Cat sample are not strictly applicable to our analysis due to the heterogeneous nature of the extended catalogue, the fact that a similar effect is seen for the considerably larger sample with 862 in the 1--5~Gyr age range suggests the diminished differences cannot be explained by the reduced sample size alone.

In contrast, while the excess of HJs in overdensities is strongly diminished ($p_{\rm Poisson} \approx 0.087$) in the ZGR23 sample, a moderately significant difference ($p_{\rm KS} \approx 3.5\times10^{-3}$) between the $\tilde\rho_{\rm M,20}$ distributions of HJ and non-HJ hosts persists even after applying the 1--5~Gyr age cut. As a larger fraction of the hosts resides in this age range in the ZGR23 sample (60 per cent) compared to SWEET-Cat (44 per cent), the sample size is less severely reduced by the age cut. However, we also find giant planets ($M_{\rm p} > 50~{\rm M_\oplus}$) to be overall more prevalent in overdensities, comprising 31 per cent of all planets compared to 24 per cent in underdensities. Furthermore, while the overdensity subsample contains 95 per cent of all hot Jupiters with unambiguous environment classifications, it also contains 85 per cent of all cold Jupiters. Attributing the differences to HJ formation via flyby-induced high-eccentricity migration is therefore not straightforward. The planet populations for the age-restricted samples are shown in Figure \ref{fig:planet_properties_agecut}.

To establish whether hot Jupiters are more prevalent in phase space overdensities after accounting for the observed systematic differences in host properties, we now address three questions:
\begin{enumerate*}[label=(\roman*)]
    \item Do HJ hosts have higher phase space densities $\tilde\rho_{\rm M,20}$ than non-HJ hosts after accounting for the correlation with age?
    \item Is there a difference in the number of HJ hosts after decomposing the sample into over- and underdensities and correcting for differences in host properties? and
    \item To what extent do any demographic differences arising due to the age bias reflect underlying astrophysical effects or observational selection biases?
\end{enumerate*}

\subsection{Detrending}

\begin{figure}
\includegraphics[width=\columnwidth]{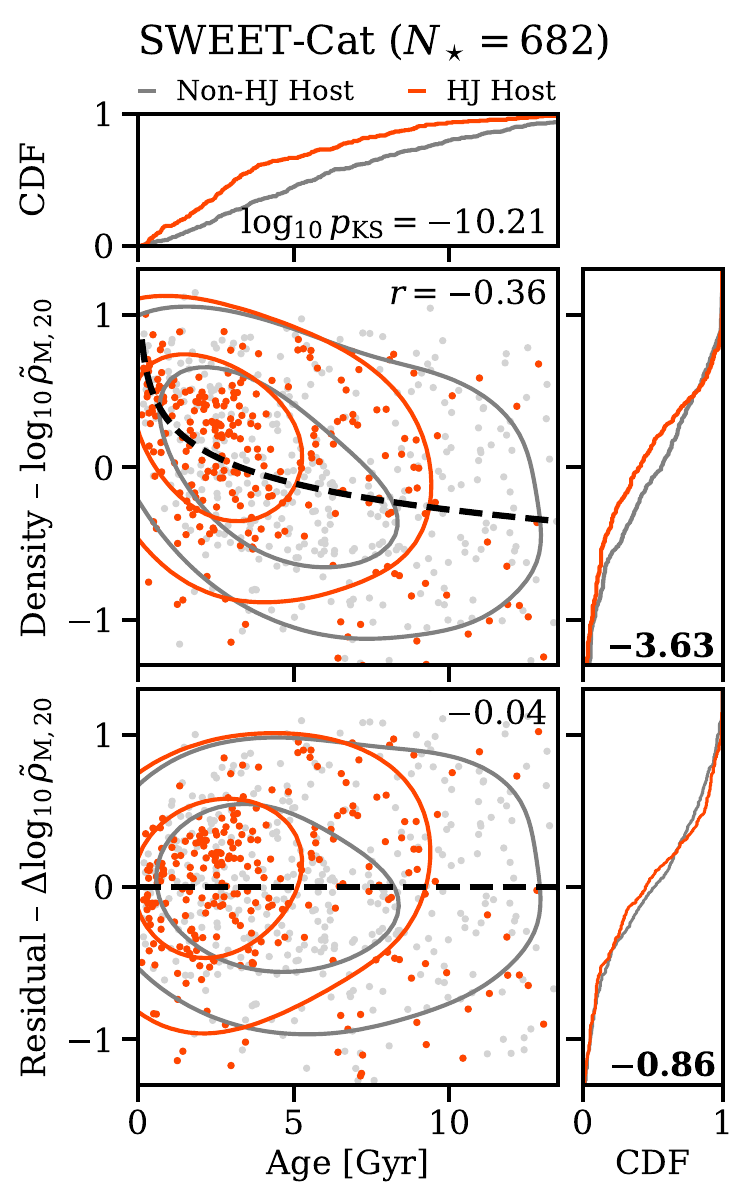}
\caption{Normalised phase space density before (middle) and residuals after (bottom) detrending against stellar age for HJ (orange) and non-HJ (grey) hosts in the homogeneous SWEET-Cat sample with no age cuts applied. The contours corresponding to the 30th and 70th percentiles of each distribution are shown. The Pearson correlation coefficient ($r$) is shown in the top right corner of each panel, and the linear regression line used for detrending is indicated with a dashed line. The marginalised cumulative distribution function of stellar age (top) shows that HJ hosts are systematically younger than non-HJ hosts. The marginal distributions and corresponding $\log_{10}$-transformed $p$-values from the two-tailed Kolmogorov-Smirnov test (right) demonstrate that HJ hosts have systematically higher phase space densities than non-HJ hosts but that this difference is substantially diminished after detrending.
\label{fig:rho_residuals_sw}}
\end{figure}

\begin{figure}
\includegraphics[width=\columnwidth]{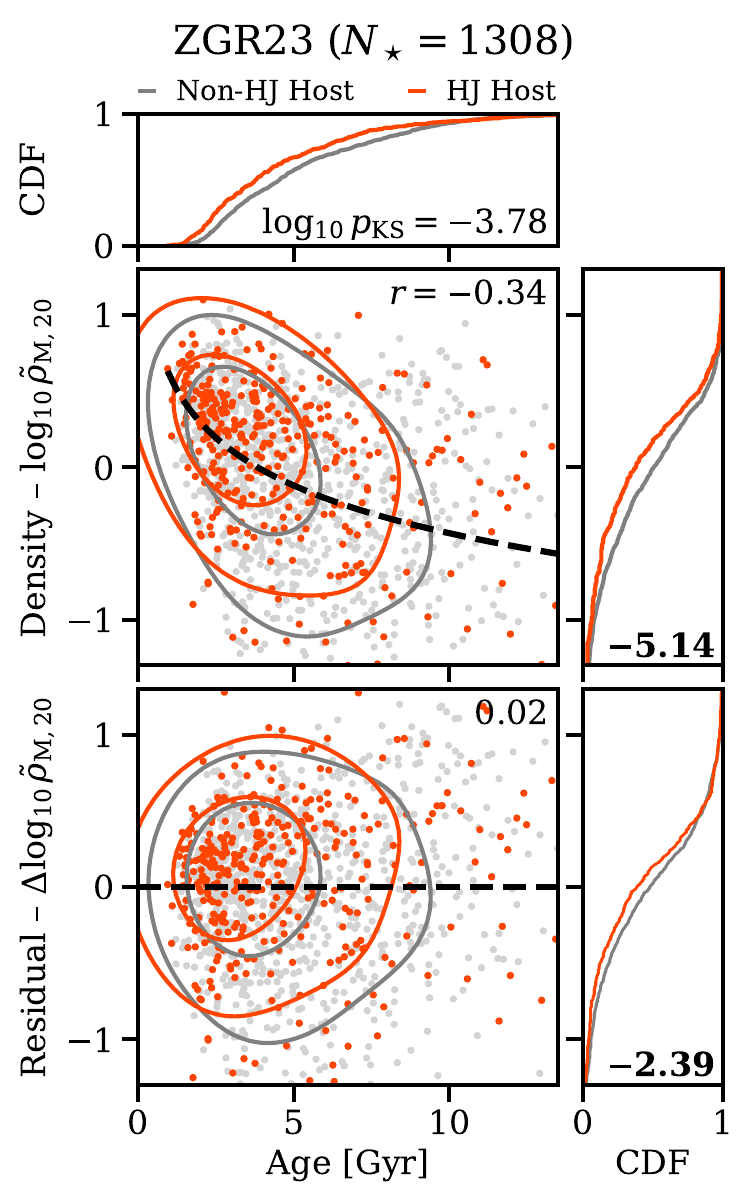}
\caption{Same as Figure \ref{fig:rho_residuals_sw} but for the ZGR23 sample. In contrast to the SWEET-Cat sample, the residual density distributions of HJ and non-HJ hosts exhibit a moderately significant difference ($\log_{10}{p_{\rm KS}} \approx -2.39$) even after detrending (bottom right).
\label{fig:rho_residuals_zgr}}
\end{figure}

\noindent To address question (i), we detrend the relationship between the normalised Mahalanobis phase space density and host age. Since velocity dispersion is generally approximated as $\sigma \propto t^\beta$ and $\log_{10}{\tilde\rho_{\rm M,20}}$ shows a strong linear anti-correlation (Pearson $r \approx -0.82$) with the total peculiar velocity $|v|$, we choose to perform the detrending against the logarithm of the host age. Specifically, we use ordinary least squares linear regression to fit a detrending line between $\log_{10}{\tilde\rho_{\rm M,20}}$ and $\log{\rm Age}$. To avoid biasing the detrending itself, we apply no age cuts prior to fitting.

The phase space density and the corresponding residuals after detrending are presented as a function of age for HJ and non-HJ hosts in the homogeneous SWEET-Cat sample in Figure \ref{fig:rho_residuals_sw}. After detrending, the phase space density residuals of HJ and non-HJ hosts show a diminished difference, with median residual $\Delta\log_{10}{\tilde\rho_{\rm M,20}} = 0.07$ and $-0.01$ for HJ and non-HJ hosts in the SWEET-Cat sample. The two distributions are no longer significantly different, with $p_{\rm KS} \approx 0.14$ as opposed to $2.4\times10^{-4}$ before detrending. When restricting the sample to hosts with ages in the 1--5~Gyr range, the differences are diminished even further, with median residuals of $0.08$ and $0.03$ for HJ and non-HJ hosts, respectively, and $p_{\rm KS} \approx 0.27$. In both cases, the residual density distributions of HJ and CJ hosts as well as of giant (GP) and non-giant planet (NGP) hosts appear statistically indistinguishable. As shown in Figure \ref{fig:rho_residuals_sw_zgr}, detrending the phase space density against age also appears to diminish correlations with mass and metallicity, as quantified by the Pearson correlation coefficient ($r$), suggesting that the initially discrepant host properties between overdensities and underdensities are largely driven by differences in stellar age. We find similar results for the full SWEET-Cat sample (Figure \ref{fig:rho_residuals_swfull}).

On the other hand, while the discrepancy in the ZGR23 sample is reduced by the detrending procedure, a significant difference between the residuals of HJ and non-HJ hosts remains (Figure \ref{fig:rho_residuals_zgr}). Using the whole age range, the median residual densities for HJ and non-HJ hosts are 0.14 and 0.02, respectively, with the distributions showing a moderately significant difference ($p_{\rm KS} \approx 4.1\times10^{-3}$). Applying the 1--5~Gyr age cut modifies the residuals slightly to 0.12 and 0.04, respectively, and the difference between the distributions becomes marginally significant ($p_{\rm KS} \approx 0.042$).

However, we also find HJ and CJ hosts to have indistinguishable ($p_{\rm KS} > 0.96$) cumulative distribution functions as a function of the detrended phase space density. While this may partly be attributed to the low number of cold Jupiters in the transit-dominated ZGR23 sample, this suggests the observed discrepancy arises due to differences in the relative occurrence rate of GP and NGP hosts rather than an excess of HJs in overdensities. This difference is evident in the bottom panels of Figure \ref{fig:planet_properties} and may reflect the higher median metallicity of hosts in overdensities. We discuss potential factors contributing to this variation in giant planet occurrence further in Section \ref{section:discussion}.

Finally, if the hot-cold boundary is set at the more commonly used 0.1~AU, rather than 0.2~AU used by \citet{winter_stellar_2020}, the statistical significance of the residual density difference is diminished in both the full ($p_{\rm KS} \approx 0.019$) and the age-restricted ZGR23 sample ($p_{\rm KS} \approx 0.098$). The concurrent reduction in the total number of stars classified as HJ hosts is nominal: 375 to 345 and 250 to 228 in the full and the age-restricted sample, respectively. This highlights the importance of consistent exoplanet classification criteria across different studies. The choice of the hot-cold boundary has no notable effect on the SWEET-Cat results.

\subsection{Matching}

\begin{figure*}
\includegraphics[width=0.8\textwidth]{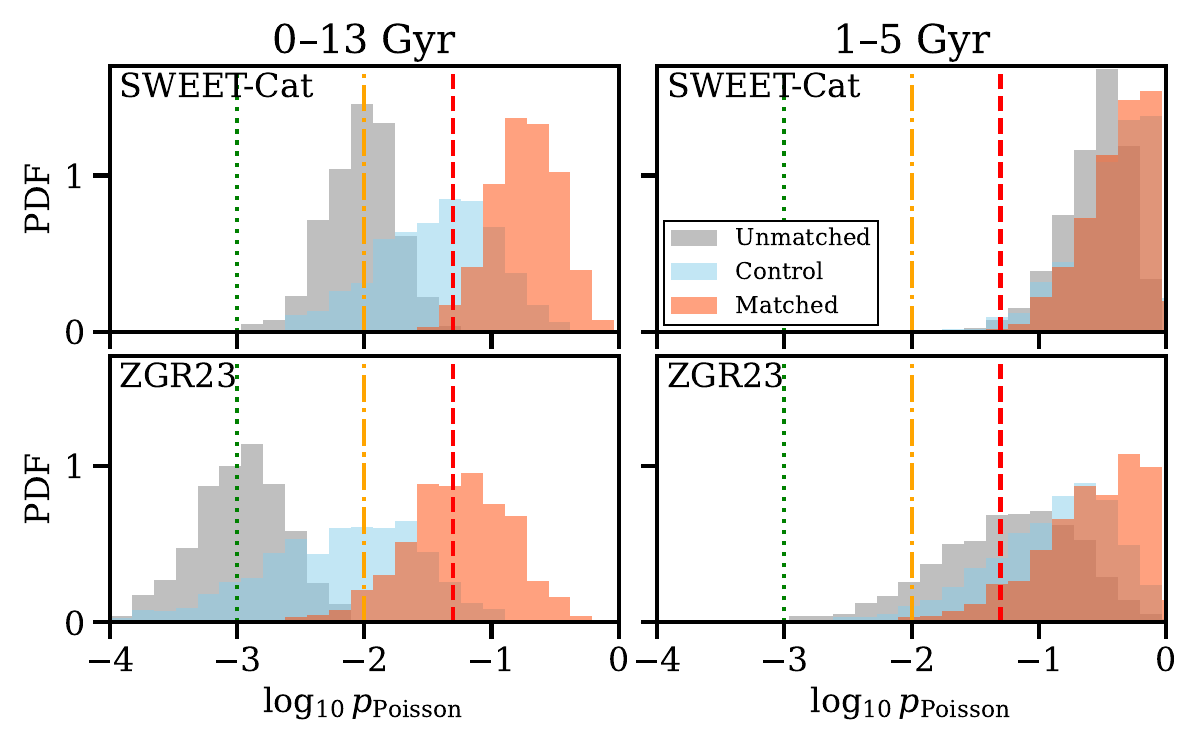}
\caption{The distribution of $p$-values from the Poisson means test quantifying the difference in the occurrence of HJ hosts between over- and underdensities using resampled age, mass, and metallicity values. The histograms represent 1000 pairs of unmatched (grey), size-controlled (blue), and like-for-like matched (orange) samples. The red, yellow, and green lines correspond to $p$-values of 0.05, 0.01, and 0.001, respectively.
\label{fig:poisson_pvals}}
\end{figure*}

\noindent Addressing question (ii) above requires that the systematic inconsistenties between hosts in the overdensity and underdensity subsamples are corrected for. To enable a direct comparison between the subsamples, we perform like-for-like matching to construct a pair of equally sized samples of host stars with similar intrinsic properties. For this purpose, the age, mass, and metallicity values are first rescaled using min-max normalisation, with each property $x$ transformed as
\begin{equation}
    x' = \frac{x-\min{x}}{\max{x}-\min{x}}
\end{equation}
such that hosts in the two subsamples can be matched across all three properties simultaneously.

For each host in the underdensity sample, we identify all potential matches in the overdensity sample based on their Euclidean distance in the rescaled feature space. To construct the final matched samples, we sort the candidate matches by increasing distance and iteratively select the closest unmatched pair, ensuring that each host is matched at most once. If no valid match is found within a predefined threshold distance, the host is excluded from the matched sample. In our experiments, we find a threshold distance of 0.25 to generally produce a good match while maximising the sample size. When comparing the planet populations in the resulting subsamples, we additionally require the age, mass, and metallicity distributions in the matched samples to be statistically consistent ($p_{\rm KS} > 0.05$). A similar procedure using $Z$-score normalisation for stellar parameter control was recently employed by \citet{tu_age_2025}. 

We first performed the matching using the nominal values of stellar age, mass, and metallicity. To account for the uncertainties in the host properties and assess the robustness of the results, we then repeated the matching procedure 1000 times, resampling the values for each host from the corresponding normal distributions based on the \texttt{isoclassify} posteriors at each iteration. The Poisson $p$-value distributions before and after the like-for-like matching procedure are presented in Figure \ref{fig:poisson_pvals}. To illustrate the impact of our matching procedure, we show the planet populations for a representative pair of matched samples in Figure \ref{fig:planet_properties_matched}.

Following the like-for-like matching, we generally find the difference in HJ occurrence between high and low phase density samples to be largely diminished and no longer statistically significant, as quantified by the Poisson means test. When performing the matching using the nominal values for age, mass, and metallicity, the resulting $p$-values are 0.28 and 0.080 in the homogeneous SWEET-Cat and ZGR23 samples, respectively. In 1000 iterations using resampled values and the full age range (0--13~Gyr), the corresponding median $p$-values are 0.19 and 0.063 with 3.1 and 39.7 per cent of the draws producing a significant result at the $p < 0.05$ level. Prior to matching, both samples show a significant difference in all 1000 iterations. We note that since the repeated hypothesis tests are performed using resampled rather than new data, the resulting $p$-values are not independent. The high fraction of significant draws produced by the ZGR23 sample therefore cannot be interpreted as grounds for rejecting the null hypothesis. As discussed above, the initial differences between the 1--5~Gyr age-restricted samples are not generally significant, and these are further diminished by the matching procedure.

As the matched sample size is limited by the smaller of the two original samples, we test whether the decreased significance can be attributed to the smaller sample size. For each pair of matched samples, we generate a control pair by randomly drawing the same number of host stars from the unmatched over- and underdensity subsamples. The Poisson $p$-value distributions from 1000 iterations of control sampling are shown in Figure \ref{fig:poisson_pvals}. Using the full age range, 57.9 and 93.2 per cent of the draws produce a significant result at the $p < 0.05$ level in the SWEET-Cat and ZGR23 samples, respectively. While there is some overlap between the $p$-value histograms for the size-controlled and like-for-like matched samples, representing iterations where matching hosts were randomly selected, the histogram for matched samples is consistently shifted towards larger $p$-values, so the diminished difference in HJ occurrence cannot be solely attributed to the reduced sample size.

\subsection{Impact of Detection Methods}

So far, we have not weighed in on the underlying cause of the observed trend in HJ occurrence. Rather, we have implicitly assumed that the age bias between the over- and underdensity subsamples drives the observed trend in HJ occurrence through astrophysical effects, namely the tendency of hot Jupiters to orbit younger stars. However, the age bias can also modify the observed demographics via observational biases, particularly the reduced detectability of planets around young, active stars in radial velocity (RV) surveys. As the homogeneous SWEET-Cat sample contains an approximately equal number of RV- and transit-detected planets (Table~\ref{table:samples}), we can compare the two subsamples to assess the extent to which the observed overdensity preference of HJ hosts is influenced by detection biases. If the excess of hot Jupiters in overdensities has an environmental origin, we would expect to see a difference regardless of detection method. While \citet{winter_stellar_2020} found a significant difference when comparing planet populations in either the RV or transit subsample, \citet{chen_evolution_2023} found hot and cold Jupiters detected by the same method (RV/transit) to have similar distributions of $P_{\rm high}$, thereby suggesting that the original observation is driven by discrepancies in the relative number of RV and transit planets in over- and underdensities.

In our unmatched SWEET-Cat sample, the underdensity subsample is dominated by RV planets (128 out of 213 planets), while the overdensity subsample contains a higher proportion of transit planets (208 out of 385 planets). We find both RV- and transit-detected HJs to be somewhat more prevalent in overdensities, although the effect is less pronounced than when considering all planets together. However, even before controlling for the systematic differences in host age, mass, and metallicity, we find the differences in HJ occurrence not to be statistically significant when considering RV and transit subsamples separately. Specifically, neither the prevalence of RV-detected HJs among all hosts with RV-detected planets nor that of transit-detected HJs among all hosts with transit-detected planets exhibits a significant difference, with Poisson $p$-values of 0.15 and 0.12, respectively. Only when considering RV and transit hosts together does the difference appear significant ($p_{\rm Poisson} \approx 3.0\times10^{-3}$).

To assess whether the lack of a significant difference in HJ occurrence within the RV and transit subsamples is simply a result of reduced sample size, we perform random sampling from the full over- and underdensity samples down to the size of the RV and transit subsamples. In 1000 such random draws, we find a median Poisson $p$ value of $\sim 0.02$ with approximately 70 per cent of the draws yielding a significant result at the $p < 0.05$ level. This suggests that the absence of a significant difference in the RV and transit subsamples is not purely a sample size effect.

The difference in HJ occurrence observed in the full sample therefore appears to largely arise from variation in the relative proportion of RV and transit planet hosts, which is itself linked to an age bias reflected in the host kinematics. This likely acts in conjunction with astrophysical effects, given the known age dependence of hot Jupiter occurrence \citep{miyazaki_evidence_2023}. We note here that, in the homogeneous SWEET-Cat sample, HJ hosts appear systematically younger than other planet-hosting stars in both the RV (3.3 vs 5.2~Gyr) and transit (3.1 vs 3.8~Gyr) subsamples. The younger age of HJ hosts is therefore not simply a result of comparing planets discovered by methods with intrinsically different detection biases but likely reflects a genuine age dependence in the hot Jupiter population.

\section{Discussion}
\label{section:discussion}

\noindent Our results suggest that the trends in planetary system properties seen in different phase space environments likely reflect underlying stellar demographic differences rather than distinct formation or evolution mechanisms. As suggested by \citet{adibekyan_stellar_2021} and \citet{mustill_hot_2022}, the differences in host star kinematics reflect an age bias, with hosts in overdensities being significantly younger. Recently, \citet{rampalli_disentangling_2025} likewise found the variation in HJ occurrence between different phase space density regimes to be attributable to a metallicity difference. Our analysis suggests the variation in host star mass and metallicity distributions is driven primarily by underlying differences in stellar age rather than by any additional physical effects.

While an age bias fully explains the SWEET-Cat trend, a slight residual difference may persist in the ZGR23 sample after detrending the phase space density against age or matching hosts in phase space over- and underdensities like-for-like. This could be due to inadequate detrending or isochronal age uncertainties. While isochrone-based stellar age estimates are reasonably reliable near the subgiant turnoff and for stars with $M_\star > 1~{\rm M_\odot}$, they can introduce large uncertainties for main-sequence field stars \citep{valle_cumulative_2013, berger_gaiakepler_2020}. For stars common to both samples, the median relative age uncertainties are 28 (SWEET-Cat) and 56 per cent (ZGR23), with a median absolute age difference of 1.5~Gyr. These uncertainties can obscure the true relationship between the 6D phase space density and host age, leading to deficient detrending. Ages derived from gyrochronology are similarly unreliable due to the tidal spin-up of HJ host stars \citep{tejada_arevalo_further_2021}. Precise asteroseismological age estimates from PLATO \citep{rauer_plato_2024}. will therefore be instrumental in assessing this relationship in the future.

Other correlated factors such as planet detection method or survey selection effects may also drive the residual differences in the ZGR23 sample. Indeed, we find stars hosting transit planets to exhibit the most significant and persistent discrepancies as quantified either by the normalised Mahalanobis phase space density or the bimodal HJ host occurrence. Detection biases inherent in transit surveys include the tendency of small planets to be preferentially detected around smaller stars due to the transit depth scaling as $(R_{\rm p}/R_\star)^2$ \citep{kipping_observational_2016}. Since our analysis did not explicitly account for stellar radius, variations in host radii between the subsamples could contribute to the residual discrepancy in HJ occurrence.

We also note that the detrending procedure does not fully eliminate correlations between the normalised phase space density and intrinsic host properties. In particular, while host ages and masses in the top and bottom halves of the residual density distribution are consistent across both samples, a significant difference in metallicity remains in the ZGR23 sample ($p_{\rm KS} \approx 0.012$). Hosts in the top half (positive residuals) appear more metal-rich, with a median [Fe/H] difference of 0.03~dex between the two halves. This could at least partially account for the higher residual phase space densities of giant planets through the well-established giant planet-metallicity correlation, consistent with the findings of \citet{rampalli_disentangling_2025}.

Since the abundance of $\alpha$-process elements is also known to correlate with planetary demographics \citep{adibekyan_overabundance_2012}, we additionally obtained estimates of [Fe/H] and [$\alpha$/Fe] inferred from \textit{Gaia} XP spectra by \citet{fallows_stellar_2024}. These measurements are available for only a subset of host stars in each sample but show good agreement with the metallicities reported in SWEET-Cat and ZGR23. Comparing the top and bottom halves of the residual density distribution, we find that the differences in [$\alpha$/Fe] distributions are highly significant in the ZGR23 sample ($p_{\rm KS} = 1.7\times10^{-5}$) but less so in SWEET-Cat ($p_{\rm KS} = 0.055$). Specifically, hosts in the bottom half of the residual density distribution (negative residuals) appear $\alpha$-enhanced relative to those in the top half (positive residuals) only in the ZGR23 sample. As [$\alpha$/Fe] is commonly used as an age proxy \citep{swastik_age_2023}, this could indicate a residual correlation between phase space density and the true stellar age, which is not fully corrected for by detrending.

Although a physical origin for the residual difference in the ZGR23 sample cannot be completely ruled out, our analysis suggests that the environmental factor is less significant than initially reported by \citet{winter_stellar_2020}. Furthermore, as the residual difference is driven by discrepancies between giant and non-giant planets, the effect cannot be directly related to enhanced HJ formation via flyby-induced high-eccentricity migration in phase space overdensities. Recently, \citet{zink_scaling_2023} found that stars with high Galactic amplitudes host fewer lower-mass planets in the Kepler sample. While no meaningful trend was identified for Jupiter-mass planets due to the small sample size, a relative excess of giant planets orbiting stars in overdensities (low Galactic amplitudes) would suggest an even stronger such trend for giant planets.

Our findings confirm that the phase space density largely serves as a proxy for stellar kinematics, as identified by \citet{mustill_hot_2022}. The interpretation of the dimensionless phase space density as a meaningful indicator of dynamical disturbances by the environment therefore remains uncertain. Using $N$-body simulations of synthetic substructured regions, \citet{blaylock-squibbs_evolution_2023} found the Mahalanobis phase space density to be a poor indicator of the initial morphology of a star-forming region over Myr timescales. However, whether it can be used to constrain the potential conatality of neighbouring stars over Gyr timescales when stellar perturbations are less frequent remains unclear. To establish this, dedicated simulations tracking the long-term evolution of phase space structures in a Galactic potential would be required.

In addition, \citet{blaylock-squibbs_no_2024} recently demonstrated that stars which retain their planets over 10~Myr of dynamical evolution end up with higher Mahalanobis phase space densities irrespective of the initial stellar density. This is due to planet-hosting stars retaining an imprint of the kinematic substructure of their birth environment, whereas stars that lose their planets due to external perturbations also lose their kinematic signatures. The differences in phase space density between surviving planet hosts and the rest of the stellar population are relatively small and decline gradually over the timescale of the simulation, reflecting the fact that the encounters responsible for planetary disruption primarily occur at early times in dense substructures. Their study also focused exclusively on planets with initial semi-major axes of 5 and 30~AU. Given that close-in planets are largely shielded from disruption by external perturbations, any demographic differences in planet occurrence between regions of low and high phase space density would be expected to be driven predominantly by planets on wide orbits.

Finally, another potential confounding factor is the differential completeness of HJ and cold Jupiter (CJ) host neighbourhoods in \textit{Gaia} DR3. We find HJ hosts to have significantly fewer stellar neighbours within a 40~pc radius compared to CJ hosts, with median counts of 2,600 and 11,400, respectively. These differences are largely due to the astrometric completeness diminishing strongly with distance, which may introduce systematic differences in the inferred phase space properties between the relatively nearby RV planet hosts and the more distant transit planet hosts. While we find no clear correlation between the number of neighbours and the Mahalanobis phase space density, the potential contribution of this difference to the observed variation in the relative number of RV and transit planet hosts assigned to over- and underdensities in the SWEET-Cat sample cannot be ruled out.

\section{Conclusions}
\label{section:conclusions}

\noindent In this work we have studied the observed correlation between hot Jupiter occurrence and Mahalanobis phase space density using two homogeneous and significantly larger host samples than those considered in previous studies. We have found that the apparent excess of hot Jupiters orbiting overdensity stars likely results from established correlations between kinematics and host star properties, particularly age. The properties of planetary systems associated with host stars in phase space overdensities and underdensities are significantly different before discrepancies in host age, mass, and metallicity are taken into account. After correcting for these either by detrending or matching hosts like-for-like, the differences in planetary system properties are strongly diminished and are at most marginally significant.

Our results demonstrate the importance of comparing samples that are not only statistically consistent in terms of individual host properties but also exhibit consistent multivariate distributions. We note that our work pertains only to the use of the Mahalanobis phase space density as a proxy for stellar clustering and does not rule out the possibility of other correlations between planetary properties and the stellar environment, leaving room for further investigation with larger and more comprehensive datasets.

\section*{Acknowledgements}
We thank the anonymous referee for their helpful comments which improved the clarity of this manuscript. MVK thanks the UK Science and Technology Facilities Council (STFC) for a Ph.D. studentship and also thanks Ted von Hippel and James Rogers for useful discussions. CJC has been supported by the UK Science and Technology research Council (STFC) via the consolidated grant ST/W000997/1. AJW is supported by the Royal Society through a University Research Fellowship, grant number URF\textbackslash R1\textbackslash 241791.

\section*{Data Availability}

The data used and generated in this study are publicly available on Zenodo at \href{https://doi.org/10.5281/zenodo.15865437}{https://doi.org/10.5281/zenodo.15865437}.

\bibliographystyle{mnras}
\bibliography{references}

\begin{thebibliography}{}
\makeatletter
\relax
\def\mn@urlcharsother{\let\do\@makeother \do\$\do\&\do\#\do\^\do\_\do\%\do\~}
\def\mn@doi{\begingroup\mn@urlcharsother \@ifnextchar [ {\mn@doi@} {\mn@doi@[]}}
\def\mn@doi@[#1]#2{\def\@tempa{#1}\ifx\@tempa\@empty \href {http://dx.doi.org/#2} {doi:#2}\else \href {http://dx.doi.org/#2} {#1}\fi \endgroup}
\def\mn@eprint#1#2{\mn@eprint@#1:#2::\@nil}
\def\mn@eprint@arXiv#1{\href {http://arxiv.org/abs/#1} {{\tt arXiv:#1}}}
\def\mn@eprint@dblp#1{\href {http://dblp.uni-trier.de/rec/bibtex/#1.xml} {dblp:#1}}
\def\mn@eprint@#1:#2:#3:#4\@nil{\def\@tempa {#1}\def\@tempb {#2}\def\@tempc {#3}\ifx \@tempc \@empty \let \@tempc \@tempb \let \@tempb \@tempa \fi \ifx \@tempb \@empty \def\@tempb {arXiv}\fi \@ifundefined {mn@eprint@\@tempb}{\@tempb:\@tempc}{\expandafter \expandafter \csname mn@eprint@\@tempb\endcsname \expandafter{\@tempc}}}

\bibitem[\protect\citeauthoryear{Adibekyan et~al.,}{Adibekyan et~al.}{2012}]{adibekyan_overabundance_2012}
Adibekyan V.~Z.,  et~al., 2012, \mn@doi [A\&A] {10.1051/0004-6361/201219564}, 543, A89

\bibitem[\protect\citeauthoryear{Adibekyan et~al.,}{Adibekyan et~al.}{2021}]{adibekyan_stellar_2021}
Adibekyan V.,  et~al., 2021, \mn@doi [A\&A] {10.1051/0004-6361/202040201}, 649, A111

\bibitem[\protect\citeauthoryear{Banerjee, Narang, Manoj, Henning, Tyagi, Surya, Nayak  \& Tripathi}{Banerjee et~al.}{2024}]{banerjee_host-star_2024}
Banerjee B.,  Narang M.,  Manoj P.,  Henning T.,  Tyagi H.,  Surya A.,  Nayak P.~K.,   Tripathi M.,  2024, \mn@doi [AJ] {10.3847/1538-3881/ad429f}, 168, 7

\bibitem[\protect\citeauthoryear{Bashi \& Zucker}{Bashi \& Zucker}{2022}]{bashi_exoplanets_2022}
Bashi D.,  Zucker S.,  2022, \mn@doi [MNRAS] {10.1093/mnras/stab3596}, 510, 3449

\bibitem[\protect\citeauthoryear{Bensby, Feltzing  \& Oey}{Bensby et~al.}{2014}]{bensby_exploring_2014}
Bensby T.,  Feltzing S.,   Oey M.~S.,  2014, \mn@doi [A\&A] {10.1051/0004-6361/201322631}, 562, A71

\bibitem[\protect\citeauthoryear{Berger, Huber, Gaidos, van Saders  \& Weiss}{Berger et~al.}{2020}]{berger_gaiakepler_2020}
Berger T.~A.,  Huber D.,  Gaidos E.,  van Saders J.~L.,   Weiss L.~M.,  2020, \mn@doi [AJ] {10.3847/1538-3881/aba18a}, 160, 108

\bibitem[\protect\citeauthoryear{Berger, Schlieder  \& Huber}{Berger et~al.}{2023}]{Berger_gaia-kepler-tess-host_2023}
Berger T.~A.,  Schlieder J.~E.,   Huber D.,  2023, The {Gaia}-{Kepler}-{TESS}-{Host} {Stellar} {Properties} {Catalog}: {Uniform} {Physical} {Parameters} for 7993 {Host} {Stars} and 9324 {Planets}, \mn@doi{10.48550/arXiv.2301.11338}, \url {http://arxiv.org/abs/2301.11338}

\bibitem[\protect\citeauthoryear{Bitsch, Trifonov  \& Izidoro}{Bitsch et~al.}{2020}]{bitsch_eccentricity_2020}
Bitsch B.,  Trifonov T.,   Izidoro A.,  2020, \mn@doi [A\&A] {10.1051/0004-6361/202038856}, 643, A66

\bibitem[\protect\citeauthoryear{Blaylock-Squibbs \& Parker}{Blaylock-Squibbs \& Parker}{2023}]{blaylock-squibbs_evolution_2023}
Blaylock-Squibbs G.~A.,  Parker R.~J.,  2023, \mn@doi [MNRAS] {10.1093/mnras/stac3683}, 519, 3643

\bibitem[\protect\citeauthoryear{Blaylock-Squibbs, Parker  \& Daffern-Powell}{Blaylock-Squibbs et~al.}{2024}]{blaylock-squibbs_no_2024}
Blaylock-Squibbs G.~A.,  Parker R.~J.,   Daffern-Powell E.~C.,  2024, \mn@doi [ApJ] {10.3847/1538-4357/ad4be0}, 968, 108

\bibitem[\protect\citeauthoryear{Boettner, Dayal, Trebitsch, Libeskind, Rice, Cockell  \& Tieleman}{Boettner et~al.}{2024}]{boettner_populating_2024}
Boettner C.,  Dayal P.,  Trebitsch M.,  Libeskind N.,  Rice K.,  Cockell C.,   Tieleman B.~I.,  2024, \mn@doi [A\&A] {10.1051/0004-6361/202449557}, 686, A167

\bibitem[\protect\citeauthoryear{Bovy, Rix, Green, Schlafly  \& Finkbeiner}{Bovy et~al.}{2016}]{Bovy_galactic_2016}
Bovy J.,  Rix H.-W.,  Green G.~M.,  Schlafly E.~F.,   Finkbeiner D.~P.,  2016, \mn@doi [ApJ] {10.3847/0004-637X/818/2/130}, 818, 130

\bibitem[\protect\citeauthoryear{Brucalassi et~al.,}{Brucalassi et~al.}{2016}]{brucalassi_search_2016}
Brucalassi A.,  et~al., 2016, \mn@doi [A\&A] {10.1051/0004-6361/201527561}, 592, L1

\bibitem[\protect\citeauthoryear{Chen et~al.,}{Chen et~al.}{2023}]{chen_evolution_2023}
Chen D.-C.,  et~al., 2023, \mn@doi [PNAS] {10.1073/pnas.2304179120}, 120, e2304179120

\bibitem[\protect\citeauthoryear{Choi, Dotter, Conroy, Cantiello, Paxton  \& Johnson}{Choi et~al.}{2016}]{Choi_mesa_2016}
Choi J.,  Dotter A.,  Conroy C.,  Cantiello M.,  Paxton B.,   Johnson B.~D.,  2016, \mn@doi [ApJ] {10.3847/0004-637X/823/2/102}, 823, 102

\bibitem[\protect\citeauthoryear{Dawson \& Johnson}{Dawson \& Johnson}{2018}]{dawson_origins_2018}
Dawson R.~I.,  Johnson J.~A.,  2018, \mn@doi [ARA\&A] {10.1146/annurev-astro-081817-051853}, 56, 175

\bibitem[\protect\citeauthoryear{Dotter}{Dotter}{2016}]{Dotter_mesa_2016}
Dotter A.,  2016, \mn@doi [ApJS] {10.3847/0067-0049/222/1/8}, 222, 8

\bibitem[\protect\citeauthoryear{Drimmel, Cabrera-Lavers  \& López-Corredoira}{Drimmel et~al.}{2003}]{Drimmel_three-dimensional_2003}
Drimmel R.,  Cabrera-Lavers A.,   López-Corredoira M.,  2003, \mn@doi [A\&A] {10.1051/0004-6361:20031070}, 409, 205

\bibitem[\protect\citeauthoryear{El-Badry, Rix  \& Heintz}{El-Badry et~al.}{2021}]{el-badry_million_2021}
El-Badry K.,  Rix H.-W.,   Heintz T.~M.,  2021, \mn@doi [MNRAS] {10.1093/mnras/stab323}, 506, 2269

\bibitem[\protect\citeauthoryear{Fallows \& Sanders}{Fallows \& Sanders}{2024}]{fallows_stellar_2024}
Fallows C.~P.,  Sanders J.~L.,  2024, \mn@doi [MNRAS] {10.1093/mnras/stae1303}, 531, 2126

\bibitem[\protect\citeauthoryear{Fischer \& Valenti}{Fischer \& Valenti}{2005}]{fischer_planet-metallicity_2005}
Fischer D.~A.,  Valenti J.,  2005, \mn@doi [ApJ] {10.1086/428383}, 622, 1102

\bibitem[\protect\citeauthoryear{Freeman}{Freeman}{1987}]{freeman_galactic_1987}
Freeman K.~C.,  1987, \mn@doi [ARA\&A] {10.1146/annurev.aa.25.090187.003131}, 25, 603

\bibitem[\protect\citeauthoryear{{Gaia Collaboration} et~al.,}{{Gaia Collaboration} et~al.}{2023}]{gaia_collaboration_gaia_2023}
{Gaia Collaboration} et~al., 2023, \mn@doi [A\&A] {10.1051/0004-6361/202243940}, 674, A1

\bibitem[\protect\citeauthoryear{Green, Schlafly, Zucker, Speagle  \& Finkbeiner}{Green et~al.}{2019}]{Green_3d_2019}
Green G.~M.,  Schlafly E.,  Zucker C.,  Speagle J.~S.,   Finkbeiner D.,  2019, \mn@doi [ApJ] {10.3847/1538-4357/ab5362}, 887, 93

\bibitem[\protect\citeauthoryear{Hamer \& Schlaufman}{Hamer \& Schlaufman}{2019}]{hamer_hot_2019}
Hamer J.~H.,  Schlaufman K.~C.,  2019, \mn@doi [AJ] {10.3847/1538-3881/ab3c56}, 158, 190

\bibitem[\protect\citeauthoryear{Helled et~al.,}{Helled et~al.}{2014}]{helled_giant_2014}
Helled R.,  et~al., 2014, in Beuther H.,  Klessen R.~S.,  Dullemond C.~P.,   Henning T.,  eds, Protostars and {Planets} {VI}. pp 643--665, \mn@doi{10.2458/azu_uapress_9780816531240-ch028}

\bibitem[\protect\citeauthoryear{Huber et~al.,}{Huber et~al.}{2017}]{Huber_asteroseismology_2017}
Huber D.,  et~al., 2017, \mn@doi [ApJ] {10.3847/1538-4357/aa75ca}, 844, 102

\bibitem[\protect\citeauthoryear{Johnson, Aller, Howard  \& Crepp}{Johnson et~al.}{2010}]{johnson_giant_2010}
Johnson J.~A.,  Aller K.~M.,  Howard A.~W.,   Crepp J.~R.,  2010, \mn@doi [PASP] {10.1086/655775}, 122, 905

\bibitem[\protect\citeauthoryear{Kamdar, Conroy, Ting, Bonaca, Johnson  \& Cargile}{Kamdar et~al.}{2019a}]{kamdar_dynamical_2019}
Kamdar H.,  Conroy C.,  Ting Y.-S.,  Bonaca A.,  Johnson B.,   Cargile P.,  2019a, \mn@doi [ApJ] {10.3847/1538-4357/ab44be}, 884, 173

\bibitem[\protect\citeauthoryear{Kamdar, Conroy, Ting, Bonaca, Smith  \& Brown}{Kamdar et~al.}{2019b}]{kamdar_stars_2019}
Kamdar H.,  Conroy C.,  Ting Y.-S.,  Bonaca A.,  Smith M.~C.,   Brown A. G.~A.,  2019b, \mn@doi [ApJL] {10.3847/2041-8213/ab4997}, 884, L42

\bibitem[\protect\citeauthoryear{Kamdar, Conroy, Ting  \& El-Badry}{Kamdar et~al.}{2021}]{kamdar_spatial_2021}
Kamdar H.,  Conroy C.,  Ting Y.-S.,   El-Badry K.,  2021, \mn@doi [ApJ] {10.3847/1538-4357/abfe5d}, 922, 49

\bibitem[\protect\citeauthoryear{Kipping \& Sandford}{Kipping \& Sandford}{2016}]{kipping_observational_2016}
Kipping D.~M.,  Sandford E.,  2016, \mn@doi [MNRAS] {10.1093/mnras/stw1926}, 463, 1323

\bibitem[\protect\citeauthoryear{Kruijssen, Longmore, Chevance, Laporte, Motylinski, Keller  \& Henshaw}{Kruijssen et~al.}{2021}]{kruijssen_not_2021}
Kruijssen J. M.~D.,  Longmore S.~N.,  Chevance M.,  Laporte C. F.~P.,  Motylinski M.,  Keller B.~W.,   Henshaw J.~D.,  2021, Not the {Birth} {Cluster}: the {Stellar} {Clustering} that {Shapes} {Planetary} {Systems} is {Generated} by {Galactic}-{Dynamical} {Perturbations}, \mn@doi{10.48550/arXiv.2109.06182}, \url {http://arxiv.org/abs/2109.06182}

\bibitem[\protect\citeauthoryear{Kunimoto \& Matthews}{Kunimoto \& Matthews}{2020}]{Kunimoto_searching_2020}
Kunimoto M.,  Matthews J.~M.,  2020, \mn@doi [AJ] {10.3847/1538-3881/ab88b0}, 159, 248

\bibitem[\protect\citeauthoryear{Marshall, Robin, Reylé, Schultheis  \& Picaud}{Marshall et~al.}{2006}]{Marshall_modelling_2006}
Marshall D.~J.,  Robin A.~C.,  Reylé C.,  Schultheis M.,   Picaud S.,  2006, \mn@doi [A\&A] {10.1051/0004-6361:20053842}, 453, 635

\bibitem[\protect\citeauthoryear{Miyazaki \& Masuda}{Miyazaki \& Masuda}{2023}]{miyazaki_evidence_2023}
Miyazaki S.,  Masuda K.,  2023, \mn@doi [AJ] {10.3847/1538-3881/acff71}, 166, 209

\bibitem[\protect\citeauthoryear{Mustill, Lambrechts  \& Davies}{Mustill et~al.}{2022}]{mustill_hot_2022}
Mustill A.~J.,  Lambrechts M.,   Davies M.~B.,  2022, \mn@doi [A\&A] {10.1051/0004-6361/202140921}, 658, A199

\bibitem[\protect\citeauthoryear{Nordström et~al.,}{Nordström et~al.}{2004}]{nordstrom_geneva-copenhagen_2004}
Nordström B.,  et~al., 2004, \mn@doi [A\&A] {10.1051/0004-6361:20035959}, 418, 989

\bibitem[\protect\citeauthoryear{Osborn \& Bayliss}{Osborn \& Bayliss}{2020}]{osborn_investigating_2020}
Osborn A.,  Bayliss D.,  2020, \mn@doi [MNRAS] {10.1093/mnras/stz3207}, 491, 4481

\bibitem[\protect\citeauthoryear{Petigura et~al.,}{Petigura et~al.}{2018}]{petigura_california-kepler_2018}
Petigura E.~A.,  et~al., 2018, \mn@doi [AJ] {10.3847/1538-3881/aaa54c}, 155, 89

\bibitem[\protect\citeauthoryear{Rampalli, Ness, Newton, Vanderburg, Buck  \& Mills}{Rampalli et~al.}{2025}]{rampalli_disentangling_2025}
Rampalli R.,  Ness M.~K.,  Newton E.~R.,  Vanderburg A.,  Buck T.,   Mills J.,  2025, Disentangling {Metallicity} {Effects} in {Hot} {Jupiter} {Occurrence} {Across} {Galactic} {Birth} {Radius} and {Phase}-{Space} {Density}, \mn@doi{10.48550/arXiv.2506.16511}, \url {http://arxiv.org/abs/2506.16511}

\bibitem[\protect\citeauthoryear{Rauer et~al.,}{Rauer et~al.}{2024}]{rauer_plato_2024}
Rauer H.,  et~al., 2024, The {PLATO} {Mission}, \mn@doi{10.48550/arXiv.2406.05447}, \url {http://arxiv.org/abs/2406.05447}

\bibitem[\protect\citeauthoryear{Rodet, Su  \& Lai}{Rodet et~al.}{2021}]{rodet_correlation_2021}
Rodet L.,  Su Y.,   Lai D.,  2021, \mn@doi [ApJ] {10.3847/1538-4357/abf8a7}, 913, 104

\bibitem[\protect\citeauthoryear{Santos et~al.,}{Santos et~al.}{2013}]{santos_sweet-cat_2013}
Santos N.~C.,  et~al., 2013, \mn@doi [A\&A] {10.1051/0004-6361/201321286}, 556, A150

\bibitem[\protect\citeauthoryear{Schönrich, Binney  \& Dehnen}{Schönrich et~al.}{2010}]{schonrich_local_2010}
Schönrich R.,  Binney J.,   Dehnen W.,  2010, \mn@doi [MNRAS] {10.1111/j.1365-2966.2010.16253.x}, 403, 1829

\bibitem[\protect\citeauthoryear{Skrutskie et~al.,}{Skrutskie et~al.}{2006}]{skrutskie_two_2006}
Skrutskie M.~F.,  et~al., 2006, \mn@doi [AJ] {10.1086/498708}, 131, 1163

\bibitem[\protect\citeauthoryear{Sousa}{Sousa}{2014}]{sousa_ares_2014}
Sousa S.~G.,  2014, in Niemczura E.,  Smalley B.,   Pych W.,  eds, , Determination of {Atmospheric} {Parameters} of {B}-, {A}-, {F}- and {G}-{Type} {Stars}: {Lectures} from the {School} of {Spectroscopic} {Data} {Analyses}.
Springer International Publishing, Cham, pp 297--310, \mn@doi{10.1007/978-3-319-06956-2_26}, \url {https://doi.org/10.1007/978-3-319-06956-2_26}

\bibitem[\protect\citeauthoryear{Sousa et~al.,}{Sousa et~al.}{2021}]{sousa_sweet-cat_2021}
Sousa S.~G.,  et~al., 2021, \mn@doi [A\&A] {10.1051/0004-6361/202141584}, 656, A53

\bibitem[\protect\citeauthoryear{Swastik, Banyal, Narang, Manoj, Sivarani, Rajaguru, Unni  \& Banerjee}{Swastik et~al.}{2022}]{Swastik_galactic_2022}
Swastik C.,  Banyal R.~K.,  Narang M.,  Manoj P.,  Sivarani T.,  Rajaguru S.~P.,  Unni A.,   Banerjee B.,  2022, \mn@doi [AJ] {10.3847/1538-3881/ac756a}, 164, 60

\bibitem[\protect\citeauthoryear{Swastik, Banyal, Narang, Unni, Banerjee, Manoj  \& Sivarani}{Swastik et~al.}{2023}]{swastik_age_2023}
Swastik C.,  Banyal R.~K.,  Narang M.,  Unni A.,  Banerjee B.,  Manoj P.,   Sivarani T.,  2023, \mn@doi [AJ] {10.3847/1538-3881/ace782}, 166, 91

\bibitem[\protect\citeauthoryear{Swastik, Banyal, Narang, Unni  \& Sivarani}{Swastik et~al.}{2024}]{swastik_age_2024}
Swastik C.,  Banyal R.~K.,  Narang M.,  Unni A.,   Sivarani T.,  2024, \mn@doi [AJ] {10.3847/1538-3881/ad40ae}, 167, 270

\bibitem[\protect\citeauthoryear{Tarricq et~al.,}{Tarricq et~al.}{2021}]{tarricq_3d_2021}
Tarricq Y.,  et~al., 2021, \mn@doi [A\&A] {10.1051/0004-6361/202039388}, 647, A19

\bibitem[\protect\citeauthoryear{Tejada~Arevalo, Winn  \& Anderson}{Tejada~Arevalo et~al.}{2021}]{tejada_arevalo_further_2021}
Tejada~Arevalo R.~A.,  Winn J.~N.,   Anderson K.~R.,  2021, \mn@doi [ApJ] {10.3847/1538-4357/ac1429}, 919, 138

\bibitem[\protect\citeauthoryear{Tu, Xie, Chen  \& Zhou}{Tu et~al.}{2025}]{tu_age_2025}
Tu P.-W.,  Xie J.-W.,  Chen D.-C.,   Zhou J.-L.,  2025, \mn@doi [Nat. Astron.] {10.1038/s41550-025-02539-1}, pp 1--12

\bibitem[\protect\citeauthoryear{Valle, Dell’Omodarme, Moroni  \& Degl’Innocenti}{Valle et~al.}{2013}]{valle_cumulative_2013}
Valle G.,  Dell’Omodarme M.,  Moroni P. G.~P.,   Degl’Innocenti S.,  2013, \mn@doi [A\&A] {10.1051/0004-6361/201220069}, 549, A50

\bibitem[\protect\citeauthoryear{Wielen}{Wielen}{1977}]{wielen_diffusion_1977}
Wielen R.,  1977, A\&A, 60, 263

\bibitem[\protect\citeauthoryear{Winter, Kruijssen, Longmore  \& Chevance}{Winter et~al.}{2020}]{winter_stellar_2020}
Winter A.~J.,  Kruijssen J. M.~D.,  Longmore S.~N.,   Chevance M.,  2020, \mn@doi [Nature] {10.1038/s41586-020-2800-0}, 586, 528

\bibitem[\protect\citeauthoryear{Zakamska \& Tremaine}{Zakamska \& Tremaine}{2004}]{zakamska_excitation_2004}
Zakamska N.~L.,  Tremaine S.,  2004, \mn@doi [AJ] {10.1086/422023}, 128, 869

\bibitem[\protect\citeauthoryear{Zhang, Green  \& Rix}{Zhang et~al.}{2023}]{zhang_parameters_2023}
Zhang X.,  Green G.~M.,   Rix H.-W.,  2023, \mn@doi [MNRAS] {10.1093/mnras/stad1941}, 524, 1855

\bibitem[\protect\citeauthoryear{Zhao, Zhao, Chu, Jing  \& Deng}{Zhao et~al.}{2012}]{zhao_lamost_2012}
Zhao G.,  Zhao Y.,  Chu Y.,  Jing Y.,   Deng L.,  2012, \mn@doi [RAA] {10.1088/1674-4527/12/7/002}, 12, 723

\bibitem[\protect\citeauthoryear{Zink et~al.,}{Zink et~al.}{2023}]{zink_scaling_2023}
Zink J.~K.,  et~al., 2023, \mn@doi [AJ] {10.3847/1538-3881/acd24c}, 165, 262

\makeatother
\end{thebibliography}

\newpage
\appendix

\section{Supplementary Figures}
\label{appendix:figures}
\FloatBarrier

\begin{figure}
\includegraphics[width=\columnwidth]{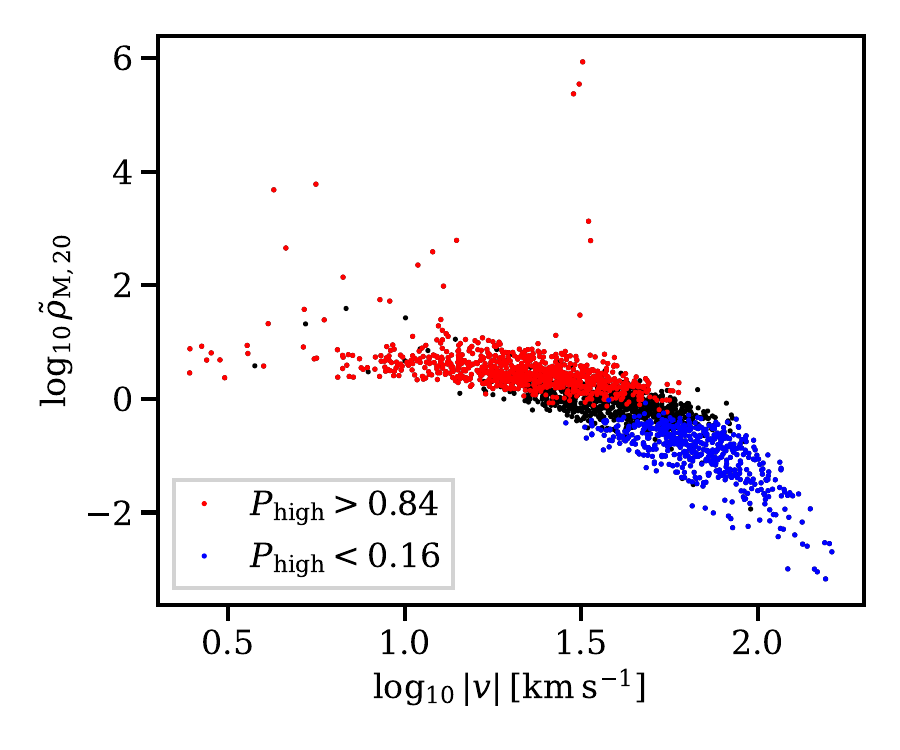}
\caption{The normalised Mahalanobis phase space density vs total peculiar velocity coloured based on division into overdensity (red), underdensity (blue), and ambiguous (black) hosts.
\label{fig:rho_vpec}}
\end{figure}

\begin{figure}
\includegraphics[width=\columnwidth]{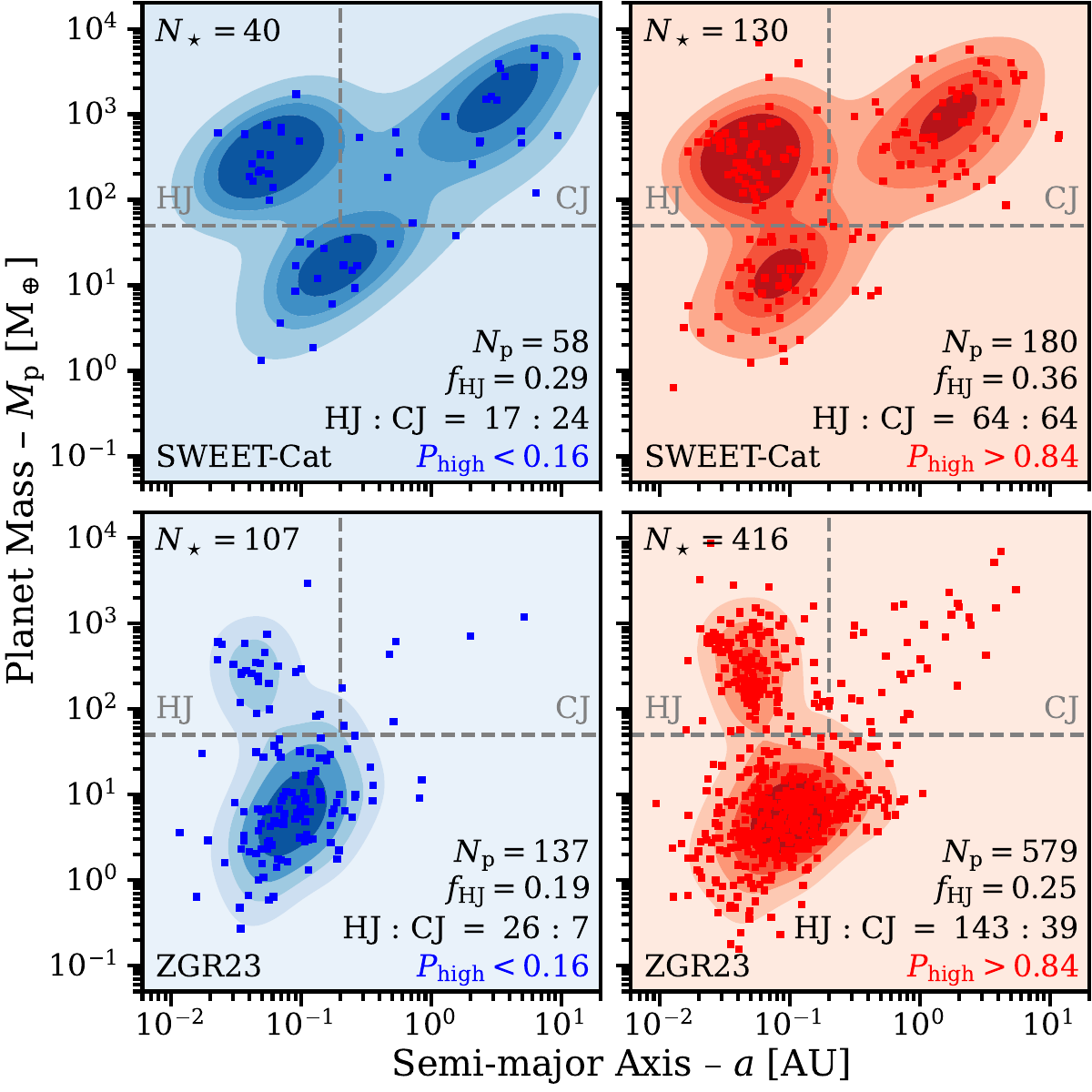}
\caption{Same as Figure \ref{fig:planet_properties} but with the 1--5~Gyr age cut applied.
\label{fig:planet_properties_agecut}}
\end{figure}

\begin{figure}
\includegraphics[width=\columnwidth]{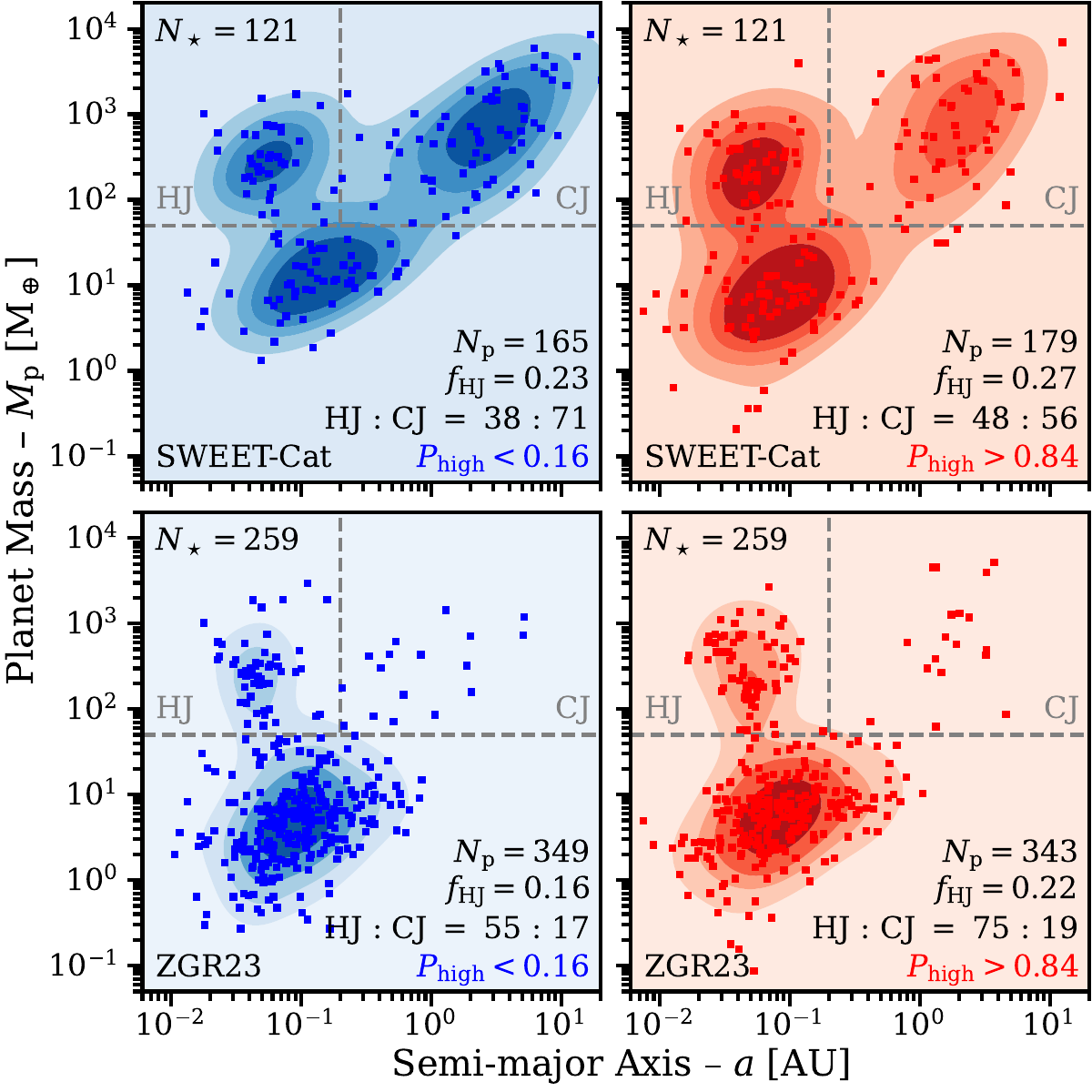}
\caption{Same as Figure \ref{fig:planet_properties} but with hosts matched like-for-like.
\label{fig:planet_properties_matched}}
\end{figure}

\begin{figure*}
\includegraphics[width=0.9\textwidth]{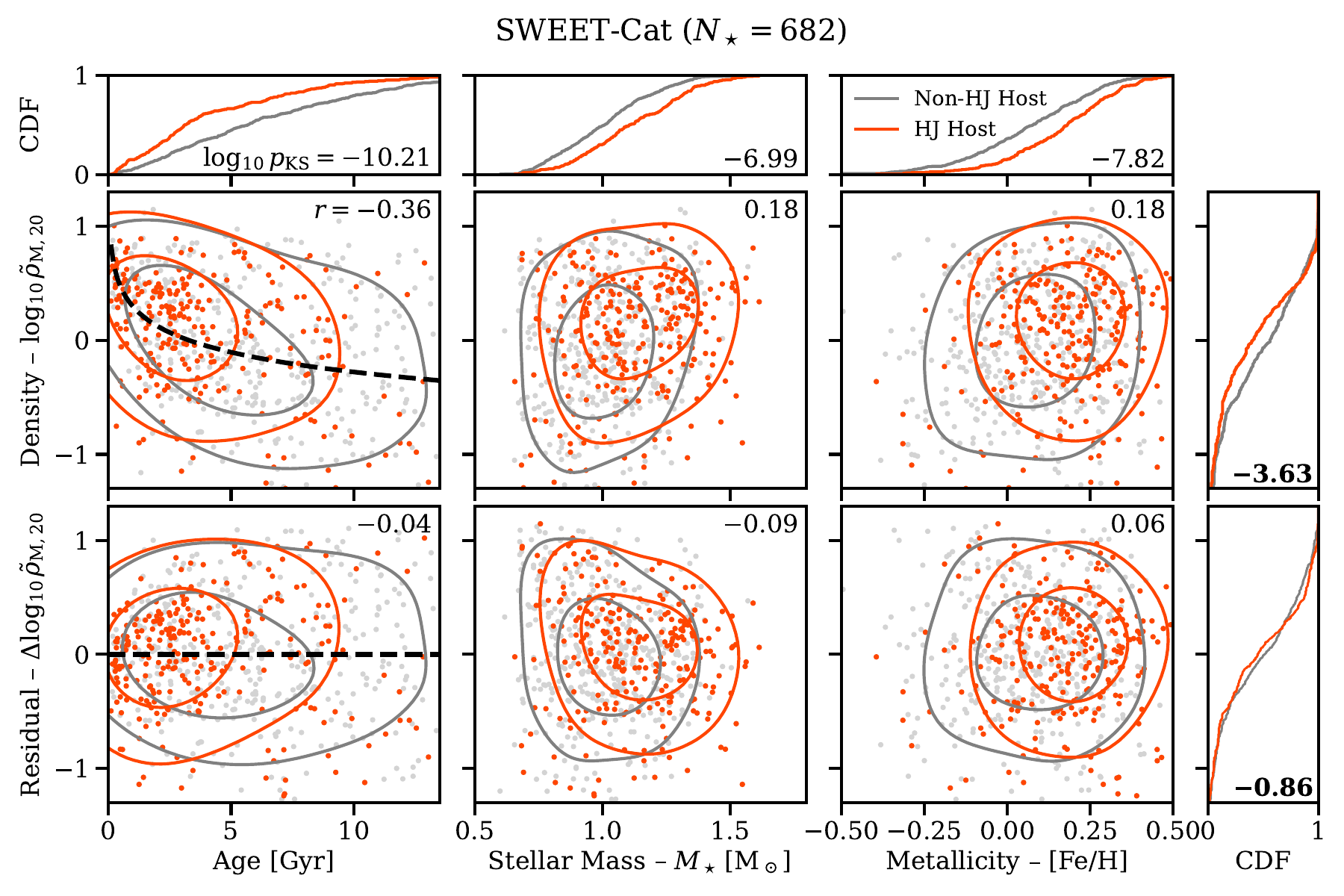}
\includegraphics[width=0.9\textwidth]{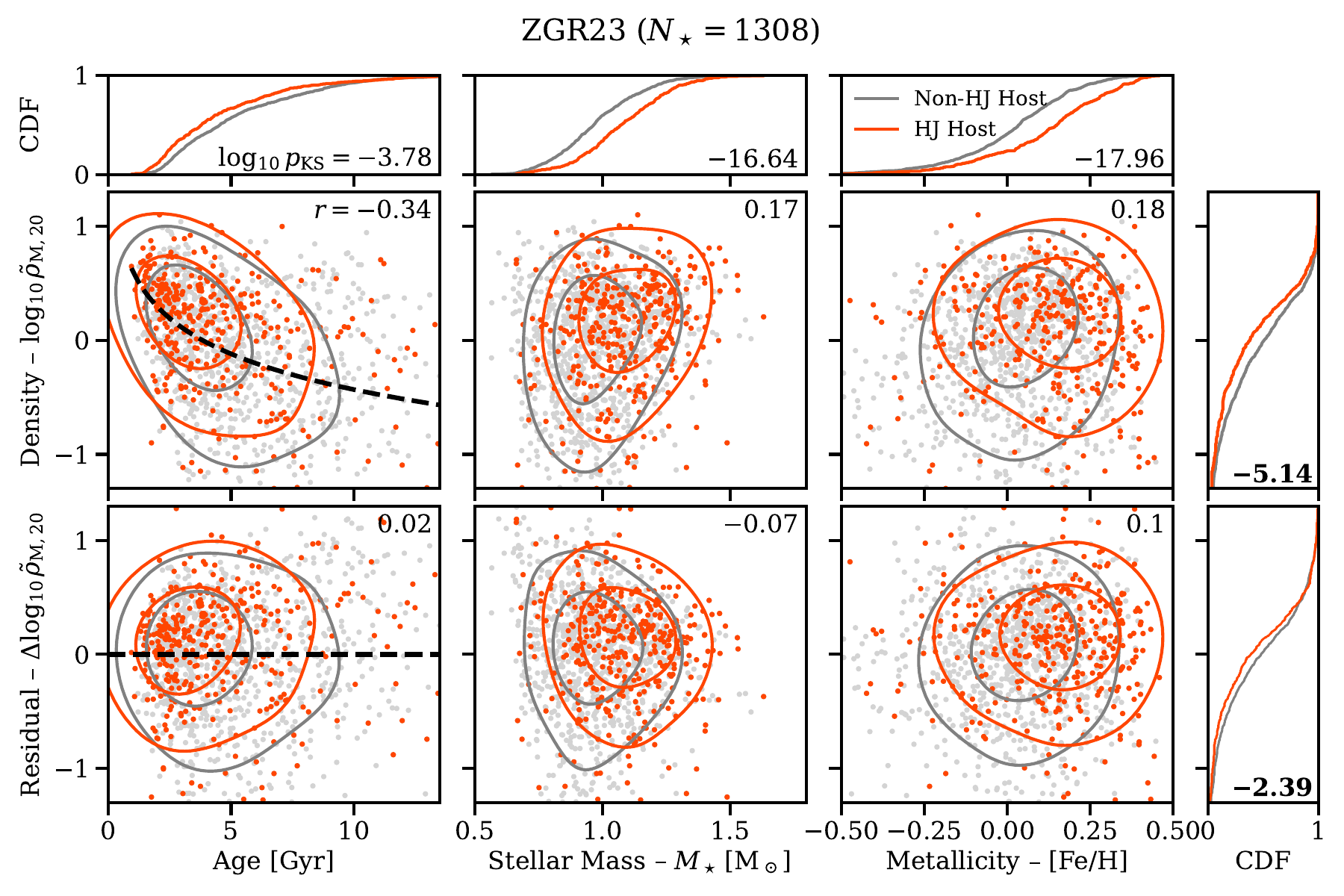}
\caption{Full summary figures of HJ (orange) and non-HJ (grey) host properties in the homogeneous SWEET-Cat (top) and ZGR23 (bottom) sample with no age cuts applied. \textit{Top row}: Cumulative distribution functions of host age, mass, and metallicity, showing that HJ hosts are systematically younger, more massive, and more metal-rich than non-HJ hosts. The corresponding $\log_{10}$-transformed $p$-value from the two-tailed Kolmogorov-Smirnov test is shown in the bottom right corner of each CDF. \textit{Middle row}: Normalised phase space density as a function of the intrinsic host properties before detrending against age. The contours corresponding to the 30th and 70th percentiles of each distribution are shown. The Pearson correlation coefficient ($r$) is shown in the top right corner of each panel, and the linear regression line used for detrending is indicated with a dashed line. The marginal distributions on the right show that HJ hosts have systematically higher phase space densities. \textit{Bottom row}: Residuals of the detrended phase space density as a function of the intrinsic host properties. The marginal distributions on the right illustrate the diminished differences between HJ and non-HJ hosts after detrending.
\label{fig:rho_residuals_sw_zgr}}
\end{figure*}

\begin{figure*}
\includegraphics[width=0.9\textwidth]{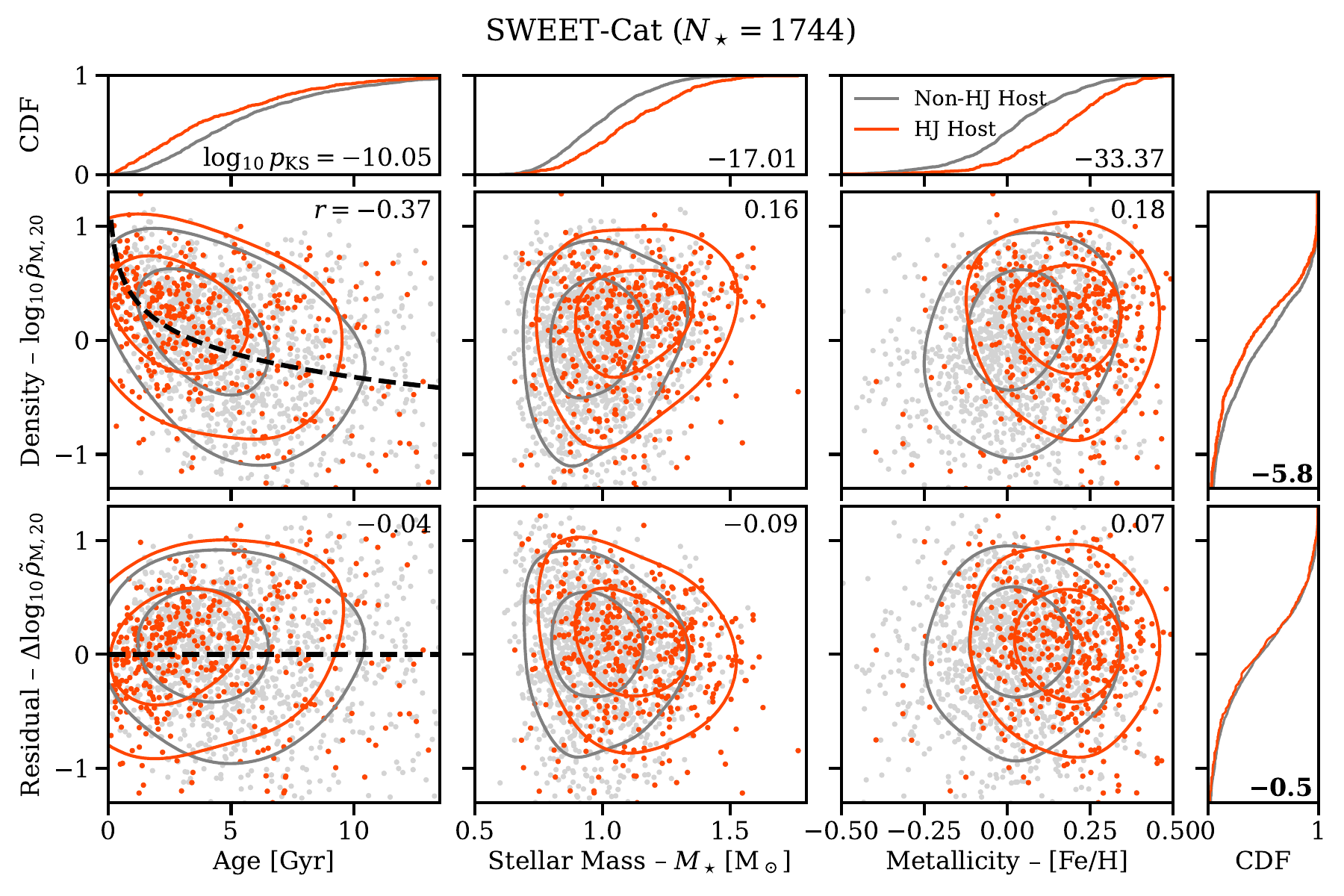}
\includegraphics[width=0.8\textwidth]{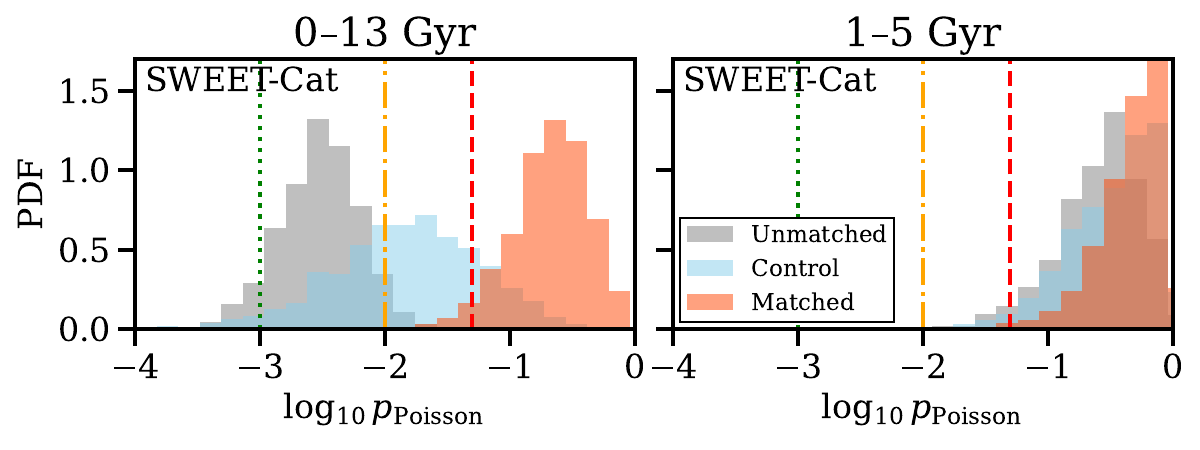}
\caption{\textit{Top}: Same as Figure \ref{fig:rho_residuals_sw_zgr} but for the full SWEET-Cat sample with no age cuts applied. \textit{Bottom}: Same as Figure \ref{fig:poisson_pvals} but for the full SWEET-Cat sample.
\label{fig:rho_residuals_swfull}}
\end{figure*}

\FloatBarrier

\bsp
\label{lastpage}
\end{document}